\documentclass[usenatbib,usegraphicx,useAMS]{mn2e}
\pdfoutput=1
\usepackage{graphicx}
\usepackage{subfig}
\usepackage{verbatim}
\usepackage[usenames,dvipsnames]{color}
\usepackage{booktabs}
\usepackage{dsfont}
\usepackage{listings} 
\usepackage{amsmath}
\usepackage{soul}
\usepackage{amssymb}

\allowdisplaybreaks[1]

\newcommand{\App}[1]{Appendix~\ref{#1}}

\newcommand{\Eq}[1]{Eq.~(\ref{#1})}

\newcommand{\Fig}[1]{Fig.~\ref{#1}}
\newcommand{\Sec}[1]{\S\ref{#1}}
\newcommand{\Tab}[1]{Table~\ref{#1}}

\def\cloudy    {{\sc Cloudy}}
\def\gadget    {{\sc Gadget}}
\def\ramses    {{\sc Ramses}}

\def\DS        {{DS12}}

\def\sbc {{\sc g9}}
\def\sbclr {{\sc g9lr}}
\def\sbclrms {{\sc g9lr\_m$_*$}}
\def\mw {{\sc g10}}
\def\dw {{\sc g8}}

\def\nh {n_{\rm{H}}}

\def\cci {{\rm{cm}}^{-3}}   
\def\kms {\rm{km} \, \rm{s}^{-1}}       
\def\msunpc {\rm M_{\odot} \, pc^{-2}}       
\def\msunyr {\rm M_{\odot} \, yr^{-1}}       
\def\Msun {\rm{M}_{\odot}}
\def\betaOut {{\beta_{\rm out}}}           

\def\chiTr {\chi_{\rm tr}}

\def\DTstoch {\Delta T_{\rm stoch}}
\def\Dx {\Delta x}
\def\Dxcen {\Delta x_{\rm cen}}
\def\Dxmax {\Delta x_{\rm max}}
\def\Dxmin {\Delta x_{\rm min}}

\def\eSN {E_{\rm SN}}
\def\eOne {E_{51}}
\def\etaSN {\eta_{\rm SN}}
\def\etaW {\eta_{\rm W}}
\def\eTurb {\epsilon_{\rm turb}}
\def\fe {f_{\rm e}}
\def\fgas {f_{\rm gas}}
\def\fk {f_{\rm k}}
\def\ftr {f_{\rm tr}}
\def\kb {k_{\rm B}}
\def\Lbox {L_{\rm box}}
\def\mp {m_{\rm p}}
\def\mcell {m_{\rm cell}}
\def\mcen {m_{\rm cen}}
\def\Mbar {M_{\rm bar}}
\def\Mbulge {M_{\rm bulge}}
\def\Mdisc {M_{\rm disc}}
\def\Mhalo {M_{\rm halo}}
\def\Mout {\dot M_{\rm out}}
\def\MSN {m_{\rm SN}}
\def\mej {m_{\rm ej}}
\def\msw {m_{\rm sw}}
\def\mW {m_{\rm W}}
\def\mstar {m_{*}}

\def\mnbor {m_{\rm nbor}}
\def\Nbor {N_{\rm nbor}}
\def\Ninj {N_{\rm inj}}
\def\Npart {N_{\rm part}}
\def\ns {n_{*}}
\def\pSN {p_{\rm SN}}

\def\rb {r_{\rm bubble}}
\def\rhonbor {\rho_{\rm nbor}}
\def\Rvir {R_{\rm vir}}

\def\sfeff {\epsilon_{*}}
\def\Siggas {\Sigma_{\rm gas}}

\def\Sigsfr {\Sigma_{\rm SFR}}
\def\SigOut {\Sigma_{\rm out}}

\def\sTurb {\sigma_{\rm turb}}
\def\tcool {t_{\rm{c}}}
\def\tDiss {t_{\rm{delay}}}
\def\TJeans {T_{\rm{J}}}      

\def\tsound {t_{\rm{s}}}
\def\vcirc {v_{\rm circ}}       
\def\vesc {v_{\rm esc}}         
\def\vzout {\left< v_{z, \rm out} \right>} 
\def\wc {w_{\rm c}}             
\def\Zsun {Z_{\odot}}           

\mathchardef\mhyphen="2D

\long\def\symbolfootnote[#1]#2{\begingroup%
\def\thefootnote{\fnsymbol{footnote}}\footnote[#1]{#2}\endgroup}

\begin{document}

\title[SN feedback in disc galaxies]{Snap, Crackle, Pop: sub-grid
  supernova feedback in AMR simulations of disc galaxies}

\author[Rosdahl, Schaye, Dubois, Kimm, \& Teyssier] {Joakim
  Rosdahl$^{1}$\thanks{E-mail: karl-joakim.rosdahl@univ-lyon1.fr}, Joop
  Schaye$^{1}$, Yohan Dubois$^{2}$, Taysun Kimm$^{3}$,
  \newauthor and Romain Teyssier$^{4}$\\
  $^1$Leiden Observatory, Leiden University, P.O. Box 9513, 2300 RA,
  Leiden, The Netherlands \\
  $^2$Institut d'Astrophysique de Paris (UMR 7095: CNRS \& UPMC
  – Sorbonne Universit\'es), 98 bis bd Arago, F-75014 Paris, France \\
  $^3$Kavli Institute for Cosmology and Institute of Astronomy,
  Madingley Road, Cambridge CB3 0HA, UK \\
  $^4$Institute for Computational Science, University of Z\"urich,
  Winterthurerstrasse 190, CH-8057 Z\"urich, Switzerland}

\maketitle
\begin{abstract}
  We compare 5 sub-grid models for supernova (SN) feedback in adaptive
  mesh refinement (AMR) simulations of isolated dwarf and L-star disc
  galaxies with $20-40$ pc resolution. The models are thermal dump,
  stochastic thermal, ``mechanical'' (injecting energy or momentum
  depending on the resolution), kinetic, and delayed cooling
  feedback. We focus on the ability of each model to suppress star
  formation and generate outflows. Our highest-resolution runs
  marginally resolve the adiabatic phase of the feedback events, which
  correspond to 40 SN explosions, and the first three models yield
  nearly identical results, possibly indicating that kinetic and
  delayed cooling feedback converge to wrong results. At lower
  resolution all models differ, with thermal dump feedback becoming
  inefficient.  Thermal dump, stochastic, and mechanical feedback
  generate multiphase outflows with mass loading factors
  $\beta \ll 1$, which is much lower than observed. For the case of
  stochastic feedback we compare to published SPH simulations, and
  find much lower outflow rates. Kinetic feedback yields fast, hot
  outflows with $\beta\sim 1$, but only if the wind is in effect
  hydrodynamically decoupled from the disc by using a large bubble
  radius. Delayed cooling generates cold, dense and slow winds with
  $\beta> 1$, but large amounts of gas occupy regions of
  temperature-density space with short cooling times.  We conclude
  that either our resolution is too low to warrant physically
  motivated models for SN feedback, that feedback mechanisms other
  than SNe are important, or that other aspects of galaxy evolution,
  such as star formation, require better treatment.
\end{abstract}

\section{Introduction} \label{Intro.sec} In our $\Lambda$CDM Universe,
most of the mass is made up of dark matter. On large scales baryons
trace the dark matter and its gravitational potential. Baryonic gas
falls into galaxies at the centres of dark matter haloes, where it
cools radiatively and collapses to form stars. By naive gravitational
arguments, star formation should be a fast affair, consuming the gas
over local free-fall times. However, from observations we know that it
is a slow and inefficient process, taking $\sim20-100$ free-fall
times, depending on the scale under consideration
\citep[e.g.][]{Zuckerman1974, Krumholz2007,
  EvansNealJ2009}. Also, while observers have a notoriously hard time
confirming the existence of gas flowing into galaxies
\citep{Crighton2013}, they instead routinely detect oppositely
directed \emph{outflows} at velocities of hundreds of km/s \citep[see
review by][]{Veilleux2005}.

To understand the non-linear problem of galaxy formation and
evolution, theorists use cosmological simulations of dark matter,
describing the flow and collapse of baryonic star-forming gas either
with directly coupled hydrodynamics or semi-analytic models. Strong
feedback in galaxies is a vital ingredient in any model of galaxy
evolution, be it hydrodynamical or semi-analytic, that comes even
close to reproducing basic observables, such as the star formation
history of the Universe, the stellar mass function of galaxies, the
Kennicutt-Schmidt relation, rotational velocities, and outflows
\citep[e.g.][]{Vogelsberger2013, Dubois2014, Hopkins2014,
  Schaye2015, Wang2015, Somerville2015}.

In order to capture the inefficient formation of stars, the first
generation of galaxy evolution models included core collapse (or type
II) supernova (SN) feedback, where massive stars ($\ga 8 \ \Msun$) end
their short lives with explosions which inject mass, metals, and
energy into the inter-stellar medium (ISM). In early hydrodynamical
simulations, the time-integrated type II SN energy of a stellar
population, $10^{51}$ ergs per SN event, was dumped thermally into the
gas neighbouring the stellar population \citep{Katz1992}. However,
such thermal dump feedback had little impact on star formation,
resulting in an over-abundance of massive and compact galaxies. This
so-called over-cooling problem is partly numerical in nature, and a
result of low resolution both in time and space. As discussed by
\cite{DallaVecchia2012}, the energy is injected into too large a
gas mass, typically resulting in much lower temperatures than those at
work in sub-pc scale SN remnants. The relatively high cooling rates at
the typical initial temperatures attained in the remnant, of
$10^5-10^6$ K, allow a large fraction of the injected energy to be
radiated away before the gas reacts hydrodynamically, resulting in
suppressed SN blasts and hence weak feedback.  Gas cooling is,
however, also a real and physical phenomenon, and while it is
over-estimated in under-resolved simulations, a large fraction of the
energy in SN remnants may in fact be radiated away instead of being
converted into large-scale bulk motion \citep{Thornton1998}.

A number of sub-resolution SN feedback models have been developed
over the last two decades for cosmological simulations, with the
primary motivation of reproducing large-scale observables, such as the
galaxy mass function, by means of efficient feedback. The four main
classes of these empirically motivated SN feedback models are i)
kinetic feedback \citep{Navarro1993}, where a fraction of the SN
energy is injected directly as momentum, often in combination with
temporarily disabling hydrodynamical forces \citep{Springel2003},
ii) delayed cooling \citep[e.g.][]{Gerritsen1997, Stinson2006},
where radiative cooling is turned off for some time in the SN remnant,
iii) stochastic feedback \citep{DallaVecchia2012}, where the SN
energy is re-distributed in time and space into fewer but more
energetic explosions, and iv) multiphase resolution elements that
side-step unnatural `average' gas states at the resolution limit
\citep{Springel2003, Keller2014}.

In principle, a physically oriented approach to implementing SN
feedback with sub-grid models is desirable. The goal is then to
inject the SN blast as it would emerge on the smallest resolved scale,
by making use of analytic models and/or high-resolution simulations
that capture the adiabatic phase, radiative cooling, the momentum
driven phase, and the interactions between different SN
remnants. However, these base descriptions usually include simplified
assumptions about the medium surrounding the SN remnant, and fail to
capture the complex inhomogeneities that exist on unresolved scales
and can have a large impact on cooling rates. In addition, even if the
SN energy is injected more or less correctly at resolved scales, it
will generally fail to evolve realistically thereafter because the
multi-phase ISM of simulated galaxies is still at best marginally
resolved. Hence there remains a large uncertainty in how efficiently
the SN blast couples to the ISM. This translates into considerable
freedom, which requires SN feedback models to be calibrated to
reproduce a set of observations \citep[see discussion
in][]{Schaye2015}.

The most recent generation of cosmological simulations has been
relatively successful in reproducing a variety of observations, in
large part thanks to the development of subgrid models for efficient
feedback and the ability to calibrate their parameters, as well as the
inclusion of efficient active-galactic nucleus (AGN) feedback in
high-mass galaxies.  However, higher-resolution simulation works
\citep[e.g.][]{Hopkins2012b, Agertz2013} suggest that SNe alone
may not provide the strong feedback needed to produce the inefficient
star formation we observe in the Universe.

Attention has thus been turning towards complementary forms of stellar
feedback, which provide additional support to the action of
SNe. Possible additional feedback mechanisms include stellar winds
\citep[e.g.][]{Dwarkadas2007, Rogers2013, Fierlinger2016},
radiation pressure (e.g. \citealt{Haehnelt1995, Thompson2005,
  Murray2010}, but see \citealt{Rosdahl2015}), and cosmic rays
\citep[e.g.][]{Booth2013, Hanasz2013, Salem2014,
  Girichidis2016}.

None the less, SN explosions remain a powerful source of energy and
momentum in the ISM and a vital ingredient in galaxy evolution. For
the foreseeable future a sub-resolution description of them will
remain necessary in cosmological simulations and even in most feasible
studies of isolated galaxies. The true efficiency of SN feedback is
still not well known, and hence we do not know to what degree we need
to improve our SN feedback sub-resolution models versus appealing to
the aforementioned complementary physics.

Rather than introducing a new or improved sub-resolution SN feedback
model, the goal of this paper is to study existing models, using
controlled and relatively inexpensive numerical experiments of
isolated galaxy discs modelled with gravity and hydrodynamics in the
Eulerian (i.e. grid-based) code \ramses{} \citep{Teyssier2002}. We
use those simulations to assess each model's effectiveness in
suppressing star formation and generating galactic winds, the main
observational constraints we have on feedback in galaxies.

We study five subgrid prescriptions for core-collapse SN feedback in
isolated galaxy discs. We explore the `maximum' and `minimum' effects
we can get from SN feedback using these models, and consider how they
vary with galaxy mass, resolution, and feedback parameters where
applicable. The simplest of those models is the `classic'
\emph{thermal dump}, where the SN energy is simply injected into the
local volume containing the stellar population. Three additional
models we consider have been implemented and used previously in
\ramses{}. These are, in chronological order, \emph{kinetic feedback},
described in \cite{Dubois2008} and used in the Horizon-AGN
cosmological simulations \citep{Dubois2014}, \emph{delayed
  cooling}, described in \cite{Teyssier2013}, and \emph{mechanical
  feedback}, described in \cite{Kimm2014} and
\cite{Kimm2015}. In addition, for this work we have implemented
\emph{stochastic feedback} in \ramses{}, adapted from a previous
implementation in the smoothed particle hydrodynamics (SPH) code
\gadget{}, described in \citet[][henceforth
\DS{}]{DallaVecchia2012}.

\begin{table*}
  \centering
  \caption
  {Simulation initial conditions and parameters for the two disc
    galaxies modelled in this paper. The listed parameters are, from 
    left to right:
    Galaxy acronym used throughout the paper, $\vcirc$: circular
    velocity at the virial radius, $\Rvir$: halo virial radius
    (defined as the radius within which the DM density is $200$ times the
    critical density at redshift zero), $\Lbox$:
    simulation box length, $\Mhalo$: DM halo mass, $\Mdisc$: disc
    galaxy mass in baryons (stars+gas), $\fgas$: disc gas
    fraction, $\Mbulge$: stellar bulge mass, $\Npart$:
    Number of DM/stellar particles, $\mstar$: mass of
    stellar particles formed during the simulations, $\Dxmax$:
    coarsest cell resolution, $\Dxmin$: finest cell resolution,
    $Z_{\rm disc}$: disc metallicity.}
  \label{sims.tbl}
  \begin{tabular}{l|rrrrrrrrrrrr}
    \toprule
    Galaxy  & $\vcirc$ & $\Rvir$ & $\Lbox$   & $\Mhalo$ 
            & $\Mdisc$ & $\fgas$ & $\Mbulge$ & $\Npart$   
            & $\mstar$ & $\Dxmax$& $\Dxmin$  &  $Z_{\rm disc}$ \\ 
    acronym & [$\kms$]   & [kpc]   & [kpc]     & [$\Msun$]
            & [$\Msun$]&         & [$\Msun$] &   
            & [$\Msun$]& [kpc]   & [pc]      & [$\Zsun$]  \\
    \midrule
    \sbc   & $65$       & $89$    & $300$     &$10^{11}$ 
           &$3.5 \times 10^9$& $0.5$   &$3.5 \times 10^8$& $10^6$     
           & $2.0 \times 10^3$ & $2.3$   & $18$      & 0.1\\
     \mw   & $140$      & $192$   & $600$     & $10^{12}$
           &$3.5 \times 10^{10}$ & $0.3$ & $3.5 \times 10^9$ & $10^6$ 
           & $1.6 \times 10^4$ & $4.7$   & $36$ & $1.0$ \\
    \bottomrule
  \end{tabular}
\end{table*}

The layout of this paper is as follows. First, we describe the setup
of our isolated galaxy disc simulations in \Sec{simulations.sec}. We
then describe the SN feedback models in \Sec{models.sec}. In
\Sec{comparison.sec} we compare results for each of these models
using their fiducial parameters in galaxy discs of two different
masses, focusing on the suppression of star formation and the
generation of outflows. In \Sec{res_conv.sec} we compare how these
results converge with numerical resolution, both in terms of physical
scale, i.e. minimum gas cell size, and also in terms of stellar
particle mass. In Sections \ref{fb_stoch.sec} - \ref{fb_kin.sec} we
take a closer look at the stochastic, delayed cooling, and kinetic
feedback models respectively, and study how varying the free
parameters in each model affects star formation, outflows and gas
morphology. The reader can skip those sections or pick out those of
interest, without straying from the thread of the paper. We discuss
our results and implications in \Sec{Discussion.sec}, and, finally, we
conclude in \Sec{Conclusions.sec}.

\section{Simulations} \label{simulations.sec}
Before we introduce the SN feedback models compared in this paper, we
begin by describing the default setup of the simulations common to all
runs.

We run controlled experiments of two rotating isolated disc galaxies,
consisting of gas and stars, embedded in dark matter (DM) haloes. The
main difference between the two galaxies is an order of magnitude
difference in mass, both baryonic and DM. We use the AMR code
\ramses{} \citep{Teyssier2002}, which simulates the interaction of
dark matter, stellar populations and baryonic gas, via gravity,
hydrodynamics and radiative cooling. The equations of hydrodynamics
are computed using the HLLC Riemann solver \citep{Toro1994} and the
MinMod slope limiter to construct variables at cell interfaces from
the cell-centred values. We assume an adiabatic index of
$\gamma = 5/3$ to relate the pressure and internal energy, appropriate
for an ideal monatomic gas. The trajectories of collisionless DM and
stellar particles are computed using a particle-mesh solver with
cloud-in-cell interpolation (\citealt{Guillet2011}; the resolution
of the gravitational force is the same as that of the hydrodynamical
solver).

\subsection{Initial conditions}
The main parameters for the simulated galaxies and their host DM haloes
are presented in \Tab{sims.tbl}. We focus most of our analysis on the
lower-mass galaxy, which we name \sbc{}. It has a baryonic mass of
$\Mbar = \Mdisc + \Mbulge = 3.8 \times 10^9 \ \Msun$, with an initial
gas fraction of $\fgas=0.5$, and it is hosted by a DM halo of mass
$\Mhalo=10^{11} \ \Msun$.  We also compare a less detailed set of
results for feedback models in a more massive galaxy, \mw{}, similar
(though somewhat lower) in mass to our Milky-Way (MW), with
$\Mbar= 3.8 \times 10^{10} \ \Msun$, $\fgas=0.3$, and
$\Mhalo=10^{12} \ \Msun$. Each simulation is run for $250$ Myr, which
is $2.5$ orbital times (at the scale radii) for both galaxy masses,
and enough for star formation and outflows to settle to quasi-static
states.

The initial conditions are generated with the {\sc MakeDisk} code by
Volker Springel \citep[see][]{Springel2005a,Kim2014}, which has been
adapted to generate \ramses{}-readable format by Romain Teyssier and
Damien Chapon. The DM halo follows an NFW density profile
\citep{Navarro1997} with concentration parameter $c=10$ and spin
parameter $\lambda=0.04$ \citep{Maccio2008}. The dark matter in each
halo is modelled by one million collisionless particles, hence the
\sbc{} and \mw{} galaxies have DM mass resolution of $10^5$ and
$10^6 \ \Msun$, respectively. The initial disc consists of stars and
gas, both set up with density profiles which are exponential in radius
and Gaussian in height from the mid-plane (scale radii of $1.5$ kpc
for \sbc{} and $3.2$ kpc for \mw{}, and scale heights one tenth of the
scale radius in both cases). The galaxies contain stellar bulges with
masses and scale radii both one tenth that of the disc.  The initial
stellar particle number is $1.1\times 10^6$ , a million of which are
in the disc and the remainder in the bulge. The mass of the initial
stellar particles is $1.7 \times 10^3$ and $10^4 \ \Msun$ for the
\sbc{} and \mw{} galaxies, respectively, close to the masses of
stellar particles formed during the simulation runs, which are shown
in \Tab{sims.tbl}. While contributing to the dynamical evolution and
gravitational potential of the rotating galaxy disc, the initial
stellar particles do not explode as SNe. This initial lack of feedback
results in over-efficient early star formation and a subsequent strong
feedback episode which typically then suppresses the star formation to
a semi-equilibrium state within a few tens of Myr (see e.g. star
formation rate plots in \Sec{sf.sec}). To overcome this shortcoming,
future improvements should include sensible age assignments to the
initial stellar particles, which could be used to perform SN feedback
right from the start of the simulations.

The temperature of the gas discs is initialised to a uniform $T=10^4$
K and the ISM metallicity $Z_{\rm disc}$ is set to $0.1$ and $1$
$\Zsun$ for the \sbc{} and \mw{} galaxies, respectively. The
circum-galactic medium (CGM) initially consists of a homogeneous hot
and diffuse gas, with $\nh=10^{-6} \ \cci$, $T=10^6 $ K, and zero
metallicity. The cutoffs for the gas discs are chosen to minimize the
density contrast between the disc edges and the CGM. The square box
widths for the \sbc{} and \mw{} galaxies are $300$ and $600$ kpc,
respectively, and we use outflow (i.e. zero gradient) boundary
conditions on all sides.

The same initial conditions and similar simulation settings were used
in \cite{Rosdahl2015}, where we studied stellar radiation feedback
in combination with thermal dump SNe. The main differences from the
setup of the previous work, apart from not including stellar
radiation, is that here we include a homogeneous UV background, we
form stellar particles that are about a factor three more massive, and
the previous work included a bug, now fixed\footnote{Thanks to Sylvia
  Ploeckinger for finding and fixing the issue.}, in metal cooling,
where the contribution of hydrogen and helium was double-counted at
Solar metallicity\footnote{We have checked and verified that the metal
  cooling bug has a negligible effect on the results of both this and
  our previous work.}. The most significant of these changes is the
larger stellar particle mass, which boosts the efficiency of thermal
dump SN feedback in suppressing star formation and, to a lesser
extent, in generating outflows.
 
\subsection{Adaptive refinement}
Each refinement level uses half the cell width of the next coarser
level, starting at the box width at the first level. Our simulations
start at level $7$, corresponding to a coarse resolution of
$2^7=128^3$ cells, and adaptively refine up to a maximum level 14,
corresponding to an effective resolution of $16384^3$ cells. This
corresponds to an optimal physical resolution of $18$ pc and $36$ pc
in the less and more massive galaxies, respectively. Refinement is
done on mass: a cell is refined if it is below the maximum refinement
level, if its total mass (DM+stars+gas) exceeds $8 \, \mstar$
(see mass values in \Tab{sims.tbl}), or if its width exceeds a quarter
of the local Jeans length
 
\subsection{Gas thermochemistry}
The gas temperature and the non-equilibrium ionization states of
hydrogen and helium are evolved with the method presented in
\cite{{Rosdahl2013}}, which includes collisional
ionization/excitation, recombination, bremsstrahlung, di-electronic
recombination, and Compton electron scattering off cosmic microwave
background photons. We include hydrogen and helium photo-ionization
and heating of diffuse gas from a redshift zero
\cite{Faucher-Giguere2009} UV background, and enforce an exponential
damping of the UV radiation above the self-shielding density of
$\nh=10^{-2} \ \cci$.  

Above $10^4$ K, the contribution to cooling from metals is added using
\cloudy{} \citep[][version 6.02]{Ferland1998} generated tables,
assuming photoionization equilibrium with a redshift zero
\cite{Haardt1996} UV background. Below $10^4$ K, we use fine
structure cooling rates from \cite{Rosen1995}, allowing the gas to
cool radiatively to $10$ K. 

\subsection{Star formation}\label{sf_model.sec}
We use a standard star formation (SF) model which follows a Schmidt
law. In each cell where the hydrogen number density is above the star
formation threshold, $\ns = 10 \ \cci$, gas is converted into stars at
a rate $\dot \rho_{*} = \sfeff \rho / t_{\rm ff}$, where $\rho$ is the
gas (mass) density and $\sfeff$ is the star formation efficiency per
free-fall time, $t_{\rm ff} = \left[ 3 \pi/(32 G \rho) \right]^{1/2}$,
where $G$ is the gravitational constant.  Stellar populations are
represented by collisionless stellar particles that are created
stochastically using a Poissonian distribution \citep[for details
see][]{Rasera2006}, which returns the stellar particle mass as an
integer multiple of $m_*$ (see \Tab{sims.tbl}). We use $\sfeff=2\%$ in
this work \citep[e.g.][]{Krumholz2007}. In future work we will
consider how varying the details of star formation affects the
efficiency of SN feedback, but that is beyond the scope of the present
paper. The stellar particle masses are given in \Tab{sims.tbl}, and
are equal to the SF density threshold, $\ns$, times the volume of a
maximally refined gas cell\footnote{We do not allow more than $90 \%$
  of the cell gas to be removed when forming stars. Thus, stellar
  particles actually do not form below a density of $1.11 \ \ns$.}.

\subsection{Artificial Jeans pressure} 
To prevent numerical fragmentation of gas below the Jeans scale
\citep{Truelove1997}, an artificial `Jeans pressure' is maintained
in each gas cell in addition to the thermal pressure. In terms of an
effective temperature, the floor can be written as
$\TJeans = T_0 \ \nh / \ns$, where we have set $T_0=500$ K (and $\ns$
is the aforementioned star formation threshold), to ensure that the
Jeans length is resolved by a constant minimum number of cell widths
at any density -- 7 and 3.5 cell widths in the smaller and larger
galaxy simulations, respectively \citep[see Eq. 3
in][]{Rosdahl2015}.  The pressure floor is non-thermal, in the
sense that the gas temperature which is evolved in the thermochemistry
is the difference between the total temperature and the floor --
therefore we can have $T << T_{\rm J}$.

\section{SN feedback} \label{models.sec} Supernova feedback is
performed with single and instantaneous injections of the cumulative
SN energy per stellar population particle. Each stellar particle has
an energy and mass injection budget of
\begin{align}
\eSN &= 10^{51} \ {\rm{erg}} \ \ \etaSN \ \frac{\mstar}{\MSN}, 
       \label{Esn.eq} \\
\mej &=\etaSN \ m_*, \label{msn.eq}
\end{align}
respectively, where $\etaSN$ is the fraction of stellar mass that is
recycled into SN ejecta\footnote{Note that we will neglect the mass
  that ends up in stellar remnants of SNe.}, $\MSN$ is the average
stellar mass of a type II SN progenitor, and, as a reminder, $\mstar$
is the mass of the stellar particle.  We assume a
\cite{Chabrier2003} initial mass function (IMF) and set
$\etaSN=0.2$ and $\MSN=10 \ \Msun$, giving at least
$40 \times 10^{51}$ ergs per particle in the \sbc{} galaxy and
$320 \times 10^{51}$ ergs in the \mw{} galaxy.  We neglect the metal
yield associated with stellar populations, i.e. the stellar particles
inject no metals into the gas, and the metallicity of the gas disc
stays at roughly the initial value of $0.1$ solar (which is negligibly
diluted due to mixing with the pristine CGM). The time delay for the
SN event is 5 Myr after the birth of the stellar particle.

The model for SN energy and mass injection, and how it affects the
galaxy properties and its environment, is the topic of this paper. We
explore five different SN models, which we now describe.

\subsection{Thermal dump feedback}
This is the most simple feedback model, and one which is well known
to suffer from catastrophic radiative losses at low resolution
\citep[e.g.][]{Katz1992}. The (thermal) energy and mass of the
exploding stellar particle are dumped into the cell hosting it, and
the corresponding mass is removed from the particle. Unless the
Sedov-Taylor phase is well resolved in both space and time, the
thermal energy radiates away before it can adiabatically develop into
a shock wave. The primary aim of each of the SN models that follow is
to overcome this `overcooling' problem. 

Note that in SPH simulations, the energy in thermal dump feedback is
typically distributed over $\sim 10^2$ neighbouring gas particles,
whereas in our implementation all the energy is injected into a single
cell. Consequently, in SPH simulations with similar resolution, the
amount of gas that is heated is typically larger. This can lead to
lower temperatures and larger radiative losses in SPH, but in the case
of strong density gradients around the feedback event, it can also
enhance feedback efficiency if SPH particles with a low density
receive part of the SN energy.

\subsection{Stochastic thermal feedback}
While the other SN models described in this paper existed previously
in \ramses{} and have been described and studied individually in
previous publications, we have for this work adapted to \ramses{} the
stochastic SN feedback model presented in
\citet[][a.k.a. \DS{}]{DallaVecchia2012}, which has so far only
been used in SPH. The idea is to heat the gas in the cell hosting the
stellar particle to a temperature high enough that the cooling time is
long compared with the sound crossing time across the cell. The energy
then has the chance to do significant work on the gas before being
radiated away, and overcooling is reduced.

As argued in \DS{}, a single SN energy injection should heat the gas
enough that the ratio between the cooling time and sound crossing time
across a resolution element is $\tcool/\tsound \ga 10$. Given a local
gas density $\nh$, a physical resolution $\Dx$, and assuming cooling
is dominated by Bremsstrahlung (true for $T\ga 10^7$ K), an expression
from \DS{} (their eq. 15) can be used to derive an approximate
required temperature increase, $\DTstoch$, to enforce this minimum
ratio and thus avoid catastrophic cooling, resulting in the condition
that
\begin{align}\label{DT_min.eq}
  \DTstoch \ga 1.1 \times 10^{7} \ {\rm K} 
  \ \left( \frac{\nh}{10 \ \cci} \right)
  \ \left( \frac{\Dx}{100 \ {\rm pc}} \right).
\end{align}

Some specified time delay after the birth of the stellar population
particle ($5$ Myr in this work), it injects its total available
energy, $\eSN$, into the hosting gas cell. Since $\eSN$ may be smaller
than what is needed for the required temperature increase $\DTstoch$,
the feedback event is done stochastically, with a probability
\begin{align}\label{pSN.eq}
\begin{split}
  \pSN & = \frac{\eSN}{\Delta \epsilon \ \mcell} \\
   &= 1.6 \
  \left( \frac{\etaSN}{0.2}  \right)
  \left( \frac{\mstar}{2 \times 10^3 \ {\rm \Msun}}  \right) \\
  & \qquad {} \left( \frac{\Dx}{18 \ {\rm pc}}  \right)^{-3}
  \left( \frac{\nh}{10 \ \cci}  \right)^{-1}
  \left( \frac{\DTstoch}{10^{7.5} \ {\rm K}}  \right)^{-1}
  ,
\end{split}
\end{align}
where $\mcell$ is the gas mass of the host cell (including the SN
ejecta) and 
\begin{align}
\Delta \epsilon=\frac{\kb \DTstoch}{(\gamma-1)\mp \mu}
\end{align}
is the required specific energy, with $\kb$ the Boltzmann constant,
$\mp$ the proton mass, and $\mu$ the mean particle mass in units of
$\mp$\footnote{We use $\mu=0.6$, assuming the gas to be
  ionized.}. When a stellar particle is due to inject SN energy,
$\pSN$ is calculated via \Eq{pSN.eq}. If $\pSN\ge 1$, the available
energy is sufficient to meet the cooling time constraint and it is
simply injected into the host cell. On the other hand, if $\pSN <1$, a
random number $r$ between 0 and 1 is drawn: only if $r<p_{\rm SN}$ is
the energy $\Delta \epsilon \ \mcell$ injected into the cell,
otherwise no energy injection takes place. With this approach, the
energy averaged over the whole simulation box and sufficiently long
time-scales is close to the available SN energy budget, as we have
confirmed in our runs -- it is just distributed unevenly in space and
time in order to overcome the cooling catastrophe. Note that the mass
(and metal) yield from the stellar particles is \emph{not} subject to
the stochastic process, but is always injected into the host cell,
regardless of whether or not the energy injection takes
place\footnote{This results in slight cooling of gas in those cells
  where stellar particles inject mass but not energy.}.

We note that our stochastic feedback implementation in AMR differs
significantly from the original SPH implementation from \DS{} in two
regards. First, whereas the probability for a feedback event varies
with the local density in AMR, the event probability is a constant
over a simulation run in SPH, since each candidate for energy
injection is a gas particle and all gas particles typically have
identical mass. Second, thermal dump explosions in SPH are normally
injected into a number of (typically $\sim 50$) neighbouring gas
particles, and the objective of the stochastic feedback model is then
to reduce the number of neighbours receiving the feedback energy in
each event (in effect making it more similar to our implementation of
thermal dump feedback). However, in AMR, the energy is only released
into the gas cell hosting the exploding stellar particle, so our
stochastic feedback model redistributes feedback events such that
some stellar particles explode with boosted energy, and some not at
all.

Naively, our stochastic feedback implementation may be presumed to
re-distribute SN energy towards lower gas densities, as the
probability for each SN event scales inversely with the density via
$\mcell$. This is not the case: there is no (average) re-distribution
over densities, since the injected energy scales inversely with the
probability, and hence the average energy injected per `candidate'
feedback event, at any density, is
\begin{align}\label{pSN2.eq}
  \pSN \ \Delta \epsilon \ \mcell = \eSN.
\end{align}

We study the effects of varying the $\DTstoch$ parameter in
Sec. \Sec{fb_stoch.sec}. For the comparison of feedback models, we
use the fiducial value of $\DTstoch=10^{7.5}$ K, because: i) it is
roughly the minimum value given by \Eq{DT_min.eq} using our star
formation density threshold, ii) in our comparison of $\DTstoch$
values in Sec. \Sec{fb_stoch.sec} we find this gives a similar star
formation and Kennicutt-Schmidt relation as higher $\DTstoch$ values
(though note that higher values of $\DTstoch$ do produce stronger
outflows), and iii) it is the fiducial value used in \DS{} and in the
EAGLE simulations \citep{Schaye2015}, allowing us to qualitatively
compare the efficiency of our AMR version of the stochastic feedback
model to that of previous SPH works. Assuming a resolution of
$\Dx=18$ pc and that stellar particles typically explode close to the
star-formation density threshold of $\ns=10 \ \cci$, \Eq{DT_min.eq}
gives $\DTstoch \ga 10^6$ K, so our chosen fiducial value is well
above the estimated requirement from \DS{}\footnote{Indeed, with
  $\DTstoch =10^{7.5}$, $\nh=10 \ \cci$, and $\Dx=18$ pc, the sound
  crossing time is $\tsound\approx 2 \times 10^4$ yr (eq. 9 in \DS{})
  and the cooling time is $\tcool \approx 3.3 \times 10^6$ yr
  $\approx 165 \ \tsound$ (eq. 13 in \DS).}. Using the same values,
\Eq{pSN.eq} shows that the probability for feedback events is below
unity at densities $\nh \ga 16 \ \cci$, i.e. slightly above our
adopted star formation density threshold, for both the \sbc{} and
\mw{} galaxies we simulate (the lower resolution in \mw{} is exactly
counter-balanced by the larger stellar particle mass).

\subsection{Delayed cooling thermal feedback}
Another widely used method for overcoming the numerical overcooling
problem in galaxy formation simulations is to turn off radiative
cooling in SN heated gas for a certain amount of time. This method,
usually referred to as \emph{delayed cooling}, has been used in SPH
simulations \citep[e.g.][]{Gerritsen1999, Stinson2006,
  Governato2010}, giving both a strong suppression in star
formation and enhancement in outflows.

\cite{Teyssier2013} introduced an AMR version of this method, which
we use in this paper. Here, a specific energy tracer, $\eTurb$, is
stored on the grid in the form of a passive scalar, and typically
associated with an unresolved turbulent energy. It is advected with
the gas and decays on a time-scale $\tDiss$ as
\begin{align}
  \frac{D\eTurb}{D t} = - \frac{\eTurb}{\tDiss}.
\end{align}
For every feedback event, the SN energy of the stellar particle,
$\eSN$, is injected as thermal energy into the host cell, as in
thermal dump feedback, but at the same time it is added to the
non-thermal energy density $\rho \, \eTurb$ in the same cell. As long
as the local ``turbulent velocity'' is above a certain limit in a
given cell, $\sTurb = \sqrt{2 \eTurb} > \sigma_{\rm min}$, radiative
cooling is disabled in that location, mimicking the non-thermal nature
of turbulent energy. When the local turbulent velocity has fallen
below $\sigma_{\rm min}$, via decay, diffusion, and mixing, radiative
cooling is enabled again.

The main free parameter in the model is $\tDiss$, which determines
how quickly the turbulent energy disappears. $\sigma_{\rm min}$ is
also an adjustable parameter, but it has more or less the same effect
as $\tDiss$, so we keep it fixed at $\sigma_{\rm min}=100 \, \kms$
(corresponding to about $0.1\%$ of the injected specific energy of a
SN, or about $1/30$th of the velocity in its unloaded remnant). The
value of $\tDiss$ can be motivated by an underlying physical
mechanism, e.g. the crossing time over a few cell widths, after which
the resolved hydrodynamics should take over the unresolved advection
of energy. The appendix of \cite{Dubois2015} derives an expression
for the choice of an appropriate $\tDiss$, given the local SN rate,
density, and resolution (their eq. A8), for which our \sbc{}
simulation settings ($\sfeff=0.02$, $\etaSN=0.2$, $\nh=10 \ \cci$,
$\Dx=18$ pc) give $\tDiss \approx 1.3$ Myr.  However, in this paper we
follow the literature \citep{Teyssier2013, Roskar2014,
  Mollitor2015, Rieder2016}, and use a much larger fiducial
value of $\tDiss=10$ Myr for the \sbc{} galaxy, and $\tDiss=20$ Myr in
low-resolution versions of \sbc{} and in the \mw{} galaxy. Assuming
decay dominates over diffusion and mixing, and assuming the SN
remnants travel at $\sim 100$ ($1,000$) km/s, our fiducial $\tDiss=10$
Myr corresponds to a delay length scale of $\sim 1$ ($10$) kpc.  We
explore variations of $\tDiss$ in \Sec{fb_dc.sec} (including values
close to that derived by \citealt{Dubois2015}).

The disadvantage of delayed cooling is that while overcooling is in
part a numerical problem, radiative cooling is a real and physical
process, without which stars would not form at all. By neglecting
radiative cooling altogether, even if for a relatively short time,
delayed cooling is likely to result in over-efficient type II SN
feedback (but, at the same time, it perhaps compensates for the
neglect of other feedback processes which may be important in galaxy
evolution). In addition, delayed cooling can result in the gas
populating parts of the temperature-density diagram where the cooling
time is short, which may yield unrealistic predictions for absorption
and emission diagnostics.

\subsection{Kinetic feedback}
We use the kinetic feedback model presented in
\citet[][]{Dubois2008}.  Here, the trick to overcoming numerical
overcooling is to skip the unresolved Sedov-Taylor phase and directly
inject the expected collective result of that phase for a stellar
population, which is an expanding momentum-conserving shock wave (or
snowplow). Note, however, that the injected kinetic energy may
subsequently be converted into thermal energy if shocks develop.

SN mass and momentum is injected into gas within a bubble radius of
the exploding stellar particle. The free parameters for the method are
$\fk$, the fraction of $\eSN$ which is released in kinetic form,
$\rb$, the radius of the bubble, and $\etaW$, the sub-resolution mass
loading factor of the Sedov-Taylor phase, describing how much mass,
relative to the stellar mass, is redistributed from the cell at the
bubble centre to the bubble outskirts.

The redistributed mass consists of two components: one is the SN
ejecta, $\mej=\etaSN \mstar$, removed from the stellar particle, the
other is the swept up mass, $\msw=\etaW\mstar$, removed from the
central host cell (no more than $25\%$ of the central cell mass is
removed, hence for individual feedback events at relatively low
densities it may happen that $\msw$ is smaller than
$\etaW\mstar$). The total wind mass is thus $\mW=\mej+\msw$, which is
redistributed uniformly (i.e. uniform density) to all cells inside the
bubble.

The kinetic energy, $\fk \eSN$, is likewise distributed to the bubble
cells, but with an injected velocity (directed radially away from the
stellar particle) that increases linearly with distance from the
centre, such as to approximate the ideal Sedov-Taylor solution:
\begin{align}
  {\bf v}(\Delta m_{\rm cell}) = f_{\rm N} v_{\rm W} 
  \frac{\bf{r}_{\rm cell}}{r_{\rm bubble}},
\end{align}
where $\Delta m_{\rm cell}$ is the mass added to the cell,
$r_{\rm cell}$ is the position of the centre of the cell relative to
the stellar particle, $f_{\rm N}\sim 1$ is a bubble normalisation
constant\footnote{The normalisation constant is the volume-weighted
  average distance from the centre, for each volume element in the
  bubble. In the ideal case of infinitely small cells, the factor is
  $1.29$.} required to ensure that the total redistributed energy is
equal to $\fk \eSN$, and
\begin{align} \label{kin_v.eq}
  v_{\rm W} = \sqrt{\frac{2 \fk \eSN}{\mW}} 
  = 3,162 \ {\rm km / s} \ \sqrt{\frac{\fk}{1+\etaW/\etaSN}}
\end{align}
is the unnormalised wind velocity, where we used \Eq{Esn.eq} for the
latter equality. Note that this is the velocity of the added mass,
i.e. each cell gains momentum
\begin{align}
  \Delta {\bf p} =  {\bf v}(\Delta m_{\rm cell}) \Delta m_{\rm cell}
  \propto \sqrt{\fk \etaSN (\etaSN + \etaW)},
\end{align}
so if the mass already in the cell is substantial compared to the
added mass, the resulting velocity change can be small. The injection
is performed in the mass-weighted frame of the SN particle (with
$\mej$) and host cell (with $\msw$). The remaining thermal energy,
$(1-\fk) \eSN$, is then distributed uniformly between the bubble
cells.

In this work, we use fiducial parameters $\fk=1$, $\etaW=1$, and
$\rb=150$ pc, a size comparable to galactic super-bubbles (note that
it is also comparable to the initial scale height of the stellar and
gas disc in our simulations, which is $150$ pc and $320$ pc for the
\sbc{} and \mw{} galaxies, respectively). These values give a velocity
for the gas ejected from the central cell (from Eq. \ref{kin_v.eq}) of
$v_{\rm W} \approx 1,300$ km/s. Our choice of $\fk=1$ implies that
there have been neither radiative losses nor momentum cancellation
from the set of unresolved individual SNe inside the bubble. We
explore the effects of a smaller bubble and higher mass loading in
\Sec{fb_kin.sec}.

\subsection{Mechanical feedback}
This model was introduced to the \ramses{} code by \citet[][see also
\citealt{Kimm2015}]{Kimm2014}, and an analogue SPH scheme was earlier
described independently in \cite{Hopkins2014}. Here, momentum is
deposited into the neighbour cells of a SN hosting cell, with the
magnitude adaptively depending on whether the adiabatic phase of the
SN remnant is captured by this small bubble of cells and the mass
within it, or whether the momentum-conserving (snowplow) phase is
expected to have already developed on this scale. In the first case,
the momentum is given by energy conservation, while in the latter
case, the final momentum, which depends via the cooling efficiency on
the density and metallicity, is given by theoretical works
\citep{Blondin1998, Thornton1998}.

In a single SN injection event, SN momentum (and any excess energy) is
distributed over the nearest neighbours (sharing at least two
vertices) of the SN host cell. The number of such cells can vary,
depending on the cell refinement, but given the extreme limit where
all the neighbours are at a finer level (i.e. half the cell width) of
the SN host cell, the maximum number of neighbours is $\Nbor=48$. When
a neighbouring cell is at the same level as the host, or one level
coarser, it is given an integer weight $\wc$ corresponding to how many
of the $\Nbor$ finer level cell units it contains ($4$ if sharing a
plane with the host, $2$ if sharing a line). The SN host cell has a
weight of $\wc=4$, so the total number of cell units receiving direct
SN energy injection is $\Ninj=\Nbor+4$.

The goal is to inject into each neighbour cell a momentum $\Delta p$,
corresponding to that generated during the energy conserving phase if
that is resolved, but to let $\Delta p$ converge towards that of the
momentum-conserving snowplow phase in the limit that the energy
conserving phase is unresolved. In each SN neighbour cell, this limit
(energy vs momentum conserving) depends on the local mass-loading,
i.e. the ratio of the local wind mass, to the SN ejecta given to that
cell,
\begin{align}
  \chi \equiv \frac{\Delta\mW}{\Delta\mej},
\end{align}
with $\Delta\mej=\frac{\wc \mej}{\Ninj}$,
$\Delta\mW=\mnbor+\frac{\wc \mcen}{\Ninj}+\Delta\mej$, $\mcen$ the
mass initially contained in the SN host cell, and
$\mnbor=\wc\rhonbor\left( \Dxcen/2 \right)^3$ the initial neighbouring
gas mass, with $\rhonbor$ and $\Dxcen$ the gas neighbour cell gas
density and host cell width, respectively.

The momentum injected into each neighbour cell, radially from the
source, is
\begin{align}
  \Delta p = \frac{\wc}{\Ninj} \! \left\{ \! \! \!
  \begin{array}{lr}
    \sqrt{2 \, \chi \, \mej \, \fe \, \eSN} & {\rm if}\,  
                         \chi < \chiTr \label{dp_mech.eq},\\
    3 \! \times \! 10^{5} \, \Msun \, \frac{\rm km}{\rm s} \, 
    \eOne^{\frac{16}{17}} \, n_0^{-\frac{2}{17}} \ Z'^{-0.14} \! \! \! 
                                            & {\rm otherwise.}
  \end{array}
          \right.
\end{align} 
Here, the upper expression represents the resolved energy conserving
phase, and comes from assuming the (final) cell gas mass of
$\Delta \mW$ receives a kinetic energy $\frac{\wc \eSN}{\Ninj}$ (we
ignore $\fe$ for the moment).  The lower expression represents the
asymptotic momentum reached in the snowplow phase, with $\eOne$ is the
total SN energy (i.e. $\eSN$) in units of $10^{51}$ erg, $n_0$ the
local hydrogen number density in units of $1 \ \cci$, and
$Z'=\max \left(Z/\Zsun,0.01\right)$. The Solar metallicity form of the
expression was derived from analytic arguments, and confirmed with
numerical experiments, in \cite{Blondin1998}, and the $Z$
dependency was added in the numerical work of \cite{Thornton1998}.

The phase transition ratio, $\chiTr$, is found by equating the
snowplow expression in \Eq{dp_mech.eq} with
$\sqrt{2 \, \chiTr \, \mej \, \ftr \, \eSN}$, where $\ftr=2/3$ is the
fraction of the SN energy assumed to be in kinetic form at the
transition. This gives
\begin{align}
\begin{split}
  \chiTr =& \frac{900}{\MSN/\Msun \ \ftr} \ \eOne^{-2/17} \ n_0^{-4/17} 
            \ Z'^{-0.28} \\
  =& \, 97 \, \left( \frac{\MSN}{10 \, \Msun}\right)^{-\frac{15}{17}}
     \left( \frac{\etaSN}{0.2}\right)^{-\frac{2}{17}}
     \left( \frac{\mstar}{2 \! \times \! 10^3 \, \Msun}\right)
     ^{-\frac{2}{17}} \label{chi_tr.eq}
  \\
   & \qquad {} \left( \frac{\nh}{10\, \cci}\right)^{-\frac{4}{17}}
    \left( \frac{Z}{0.1 \, \Zsun}\right)^{-0.28},
\end{split}
\end{align} 
where we used \Eq{Esn.eq}, $\eOne=\mej/\MSN$, and normalised to
typical values for the \sbc{} galaxy in the latter equality. The
function
\begin{align}
  \fe=1-(1-\ftr) \frac{\chi-1}{\chiTr-1}
\end{align}
ensures a smooth transition between the two expressions in
\Eq{dp_mech.eq}.

If the momentum injection results in removal of total energy in a
cell, due to cancellation of velocities, the surplus energy (initial
minus final) is added to the cell in thermal form. As it has no
preferred direction, the SN host cell receives only thermal energy. In
\cite{Kimm2014} and \cite{Kimm2015}, due to wrong book-keeping
of the surplus energy, the thermal energy injection during the
adiabatic phase was overestimated (by a factor $\sim 2 - 4$) in
regions where the swept up mass is large compared to the SN ejecta (by
a similar factor), but the correct momentum and energy was used during
the snowplow phase and the adiabatic phase with little mass loading
($\chi\sim 1$). This bug has since been corrected.

If we assume that all the cells receiving the SN energy have the same
refinement level (i.e. the same cell width) and density and that the
density is at least as high as the threshold for star formation, then
the initial mass of the neighbour dominates $\Delta \mW$ and we can
get a rough estimate for the local mass loading,
\begin{align}
\begin{split} \label{chi_mech_approx.eq}
  \chi \approx & \, \frac{\mnbor}{\Delta \mej}  \\
  =& \, \frac{\Ninj}{\wc} \frac{\rho \Dx^3}{\etaSN \mstar} \\
  =& \, 60.8 \ \ \left( \frac{\wc}{4} \right)^{-1}
     \left( \frac{\nh}{10 \, \cci} \right)
     \left( \frac{\Dx}{18 {\, \rm pc}} \right)^3 \\
   & \qquad {} \qquad {} \left( \frac{\etaSN}{0.2} \right)^{-1} 
    \left( \frac{\mstar}{2 \! \times \! 10^3 \, \Msun} \right)^{-1} \\
  =& \, 0.63 \ \chiTr \ \ \left( \frac{\wc}{4} \right)^{-1}
     \left( \frac{\nh}{10 \, \cci} \right)^{\frac{21}{17}}
     \left( \frac{\Dx}{18 {\, \rm pc}} \right)^3 \\
   &  \qquad {} \qquad {}\left( \frac{\etaSN}{0.2} \right)^{-\frac{15}{17}} 
    \left( \frac{\mstar}{2 \! \times \! 10^3 \, \Msun} 
         \right)^{-\frac{15}{17}} \\
    & \qquad {} \qquad {} 
      \left( \frac{\MSN}{10 \, \Msun}\right)^{\frac{15}{17}}
      \left( \frac{Z}{0.1 \, \Zsun}\right)^{0.28}.
\end{split}
\end{align}
Here, the last equality, which comes from comparing with
\Eq{chi_tr.eq} and normalising to the \sbc{} simulation parameters,
shows that mechanical feedback events are marginally resolved, with
the momentum injection being done using the upper expression in
\Eq{dp_mech.eq} for $\nh \la 1.6 \, \ns$, but switching to the final
snowplow momentum, i.e. the lower expression in \Eq{dp_mech.eq}, for
higher gas densities. For the \mw{} galaxy, where the resolution is
lower ($\Dx=36$ pc), the stellar mass higher
($1.6 \times 10^4 \, \Msun$), and the metallicity higher ($\Zsun$),
the SN blasts are slightly worse resolved, with
$\chi \approx 1.53 \, \chiTr$ (at $\ns$) for the same
assumptions. Here the effects of lower resolution and higher
metallicity, towards worse-resolved SN blasts, are counter-weighted by
the higher stellar particle mass.

\section{SN feedback model comparison}\label{comparison.sec}

We begin by comparing all SN feedback models using the fiducial
settings. Later in this paper we will study each feedback model in
more detail and show how the results vary with the values of the free
parameters. We focus on star formation, outflows, and galaxy
morphologies. Unless stated otherwise, our analysis will be restricted
to the lower-mass \sbc{} galaxy.

\subsection{Galaxy morphologies}
\begin{figure}
  \centering
  \subfloat
  {\includegraphics[width=0.23\textwidth]
    {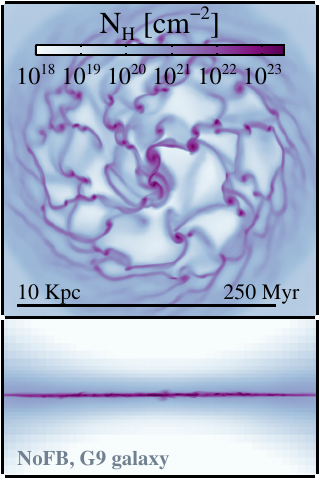}}
  \hspace{-0.6mm}
  \subfloat
  {\includegraphics[width=0.23\textwidth]
    {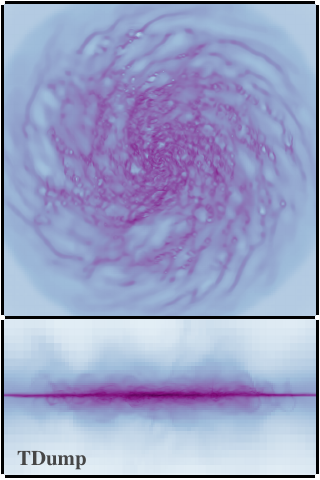}} \\
  \vspace{-3.7mm}
  \subfloat
  {\includegraphics[width=0.23\textwidth]
    {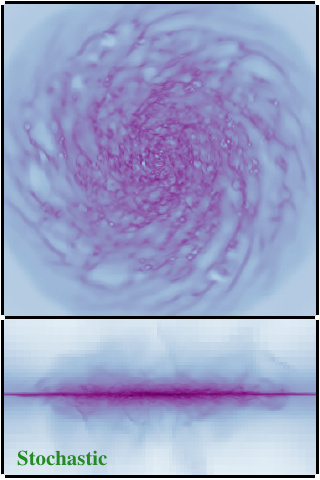}}
  \hspace{-0.6mm}
  \subfloat
  {\includegraphics[width=0.23\textwidth]
    {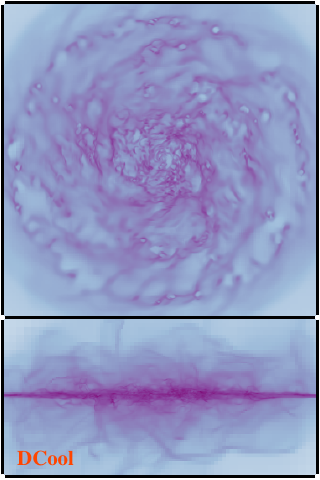}}  \\
  \vspace{-3.7mm}
  \subfloat
  {\includegraphics[width=0.23\textwidth]
    {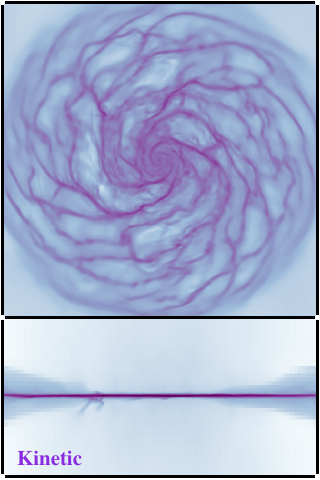}}
  \hspace{-0.6mm}
  \subfloat
  {\includegraphics[width=0.23\textwidth]
    {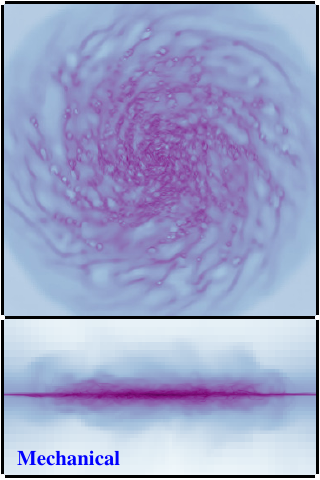}}
  \caption
  {\label{maps_Fid.fig}Maps of gas column densities in the \sbc{}
    galaxy at $250$ Myr, for the different SN feedback models. Each
    panel shows face-on and edge-on views, with the model indicated
    in the bottom left corner. The top left panel includes the
    physical length scale and the colour scale for the gas column
    density.}
\end{figure}

In \Fig{maps_Fid.fig}, we show the total hydrogen column density
face-on and edge-on at the end of the $250$ Myr run for each feedback
model. The maps illustrate how the gas morphology is affected by the
SN feedback models. Without feedback (top left panel), the galaxy
becomes clumpy, containing dense star-forming clouds which accrete
gas, thus creating large `holes'. The gas outside the thin edge-on
disc is diffuse and featureless.

Compared to the no feedback case, thermal dump feedback (top right
panel) significantly changes the gas morphology, reducing the gas
clumpiness and thickening the disc. In fact, comparing to other panels
in \Fig{maps_Fid.fig}, it has here a very similar morphological effect
as the stochastic and mechanical feedback models (middle left and
bottom right panels, respectively). We will come back to this
similarity in later sections.

Delayed cooling (middle right panel) and kinetic feedback (bottom
left), on the other hand, produce quite different morphologies from
other models in \Fig{maps_Fid.fig}. Delayed cooling diffuses the gas
more, with less obvious spiral structure, and the disc becomes
thicker, indicating increasing feedback efficiency. In stark contrast,
kinetic feedback results in a very thin disc plane, and thin, well
defined spiral filaments. Judging qualitatively from these images of
column density, delayed cooling appears most efficient in terms of
smoothing out the gas, thickening the disc, and creating outflows,
while kinetic feedback visually appears weakest, with relatively dense
and filamentary gas in the disc and low column densities out of the
disc plane. However, as we will see in what follows, kinetic feedback
actually has the strongest and fastest (but relatively diffuse)
outflows. We note that these distinct features of kinetic feedback are
sensitive to the radius of momentum and mass injection, i.e. the $\rb$
parameter. As we will argue in \Sec{SD.sec}, with our fiducial bubble
size of $150$ pc, the momentum injection is essentially
hydrodynamically decoupled from the galactic disc, and as we show in
\Sec{fb_kin.sec}, a considerably smaller bubble leads to kinetic
feedback behaving similarly to thermal dump, stochastic, and
mechanical feedback.

\begin{figure}
  \centering
  \includegraphics
    {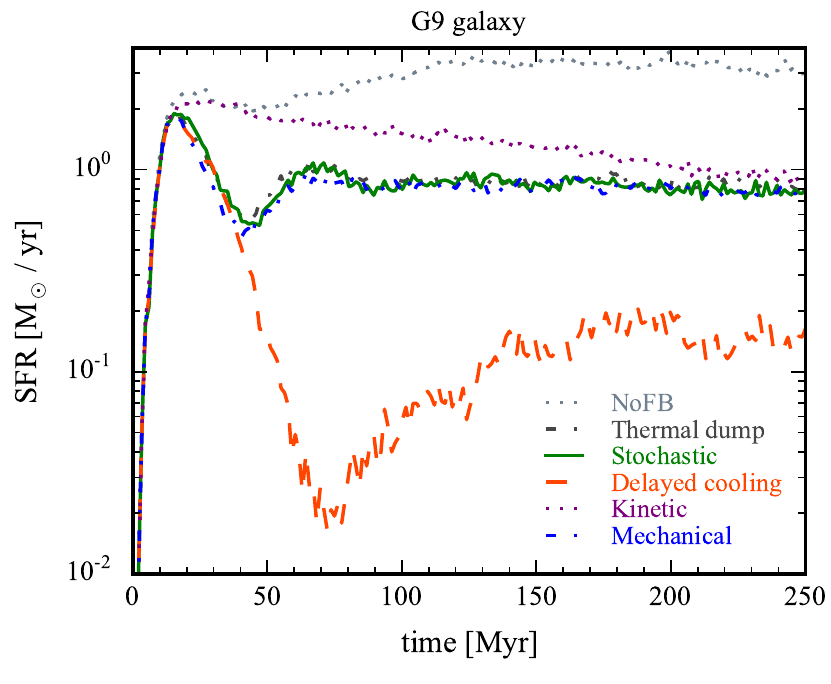}
  \caption
  {\label{SFR.fig}Star formation rates in the \sbc{} galaxy for the SN
    feedback models indicated in the legend using their fiducial
    parameters. Thermal dump, stochastic, and mechanical feedback
    produce nearly identical SFRs, while kinetic feedback produces a
    steadily declining SFR, and delayed cooling is by far the most
    efficient at suppressing star formation.}
\end{figure}
\subsection{Star formation} \label{sf.sec} The feedback efficiencies
can be quantified and compared via the star formation rates (SFRs),
which we show for the \sbc{} galaxy in \Fig{SFR.fig}. The SFRs are
calculated by binning the stellar mass formed over time intervals of
$1.2$ Myr. They vary by almost two orders of magnitude, depending on
the feedback model utilised, and one order of magnitude at the end of
the simulation runtime, by which time the rate of evolution has
settled down after the initial collapse of the disc (due to radiative
cooling and lack of initial feedback) and burst of star formation
around the $20$ Myr mark.

Focusing on the star formation around $250$ Myr, we find that the
feedback models separate roughly into the same three groups as in our
assessment of the morphologies. Thermal dump, stochastic, and
mechanical feedback all perform almost identically in terms of star
formation, indicating that thermal dump is not strongly affected by
overcooling (see \Sec{resolved.sec}). The SFR is suppressed by about a
factor of $3-4$ compared to the no feedback case (labelled NoFB in the
plot). This may seem an inefficient suppression, compared to the
inferred $1-2\%$ average efficiency of star formation observed in the
Universe, but it should be kept in mind that the star formation model
already has a built in sub-resolution efficiency of only $\sfeff=2\%$
(see \Sec{sf_model.sec}). We comment further on the choice and effect
of $\sfeff$ in the discussion (\Sec{sfe.sec}).

\begin{figure}
  \centering
  \includegraphics
    {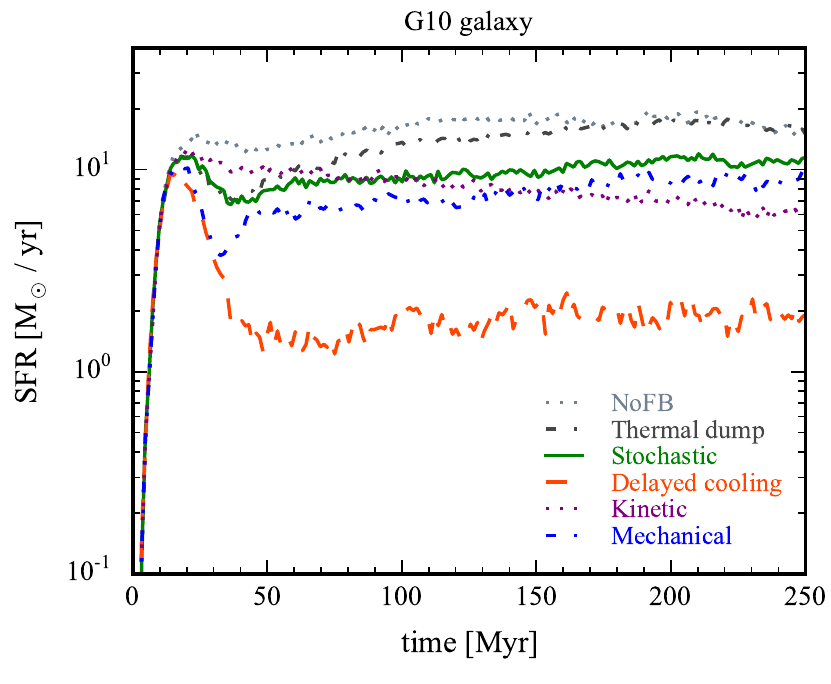}
  \caption
  {\label{SFR_MW.fig}Star formation rates, as in \Fig{SFR.fig}, but
    for the more massive and lower-resolution \mw{} galaxy (note that
    the y-axis is scaled up by a factor of ten). Here, we find
    larger differences between thermal dump, stochastic, and
    mechanical feedback, while kinetic feedback and delayed cooling
    remain qualitatively the same as in the less massive galaxy.}
\end{figure}

At $250$ Myr, kinetic feedback has a SFR fairly close to those three
aforementioned models. The difference is that the star formation rate
has not stabilised, but is declining steadily. As we will see, this is
due to the strong outflow removing gas from the star-forming ISM.

Delayed cooling is by far the most effective at suppressing star
formation. The feedback from the initial peak in the SFR almost blows
apart the gas disc, but once it has settled again the SFR stabilises
around $0.1-0.2 \, \msunyr$, though it remains somewhat bursty. The
final SFR at $250$ Myr is almost an order of magnitude lower than for
the other feedback models.

\subsubsection{Star formation in the more massive galaxy}

In \Fig{SFR_MW.fig} we show the SFRs in the ten times more massive
(and lower resolution) \mw{} galaxy simulations.

Due to the combination of the deeper gravitational potential, stronger
(metal) cooling, lower resolution, and the SN events happening at
higher gas densities (typically by $0.5-1$ dex) we find more
differences between the feedback models in their ability to suppress
star formation than for the \sbc{} galaxy. Thermal dump feedback is
weak, with the star formation stabilising at the same rate as for the
case of no feedback. With stochastic and mechanical feedback the star
formation is suppressed by about a factor of two compared to thermal
dump, with mechanical feedback being somewhat stronger.  Kinetic
feedback shows the same qualitative behaviour as in the lower-mass
galaxy, with an initially high SFR that declines steadily due to gas
outflows. Again, delayed cooling gives SFRs that are much lower than
for the other models.

\subsubsection{The Kennicutt-Schmidt relation}

\begin{figure}
  \centering
  \includegraphics
    {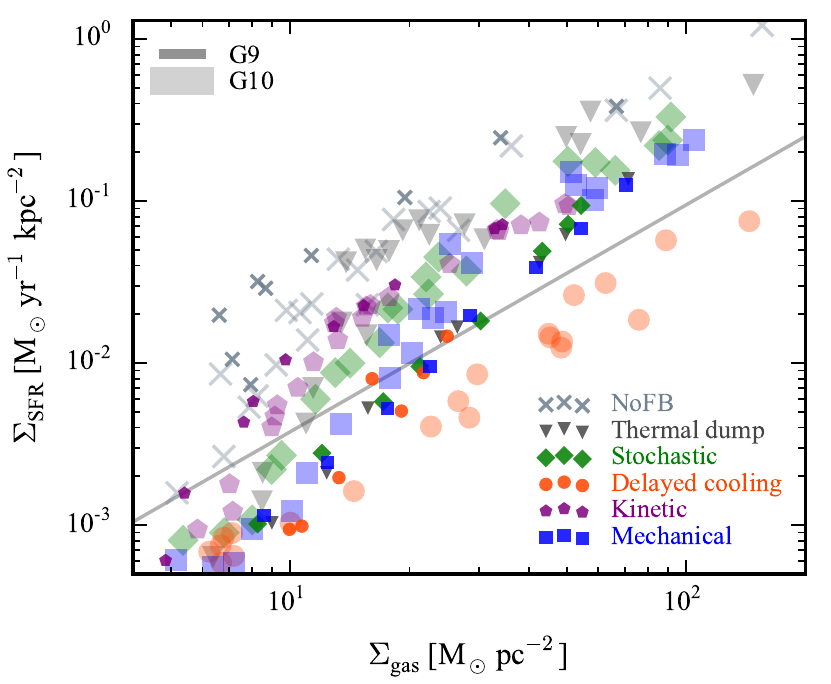}
  \caption
  {\label{KS_Fid.fig}The Kennicutt-Schmidt relation for different
    feedback models at $250$ Myr. Small opaque symbols indicate the
    \sbc{} galaxy, while larger and more transparent symbols are for
    the \mw{} galaxy. The values are averages within equally spaced
    azimuthal bins of $\Delta r=500$ pc. The grey solid line shows the
    empirical \protect\cite{Kennicutt1998} law (see text).}
\end{figure}

In the local Universe, SFR surface densities, $\Sigsfr$, are observed
on large scales to follow the Universal Kennicutt-Schmidt (KS)
relation, $\Sigsfr \propto \Siggas^{1.4}$, where $\Siggas$ is the gas
surface density \citep{Kennicutt1998}. We plot in \Fig{KS_Fid.fig}
the relation between SFR and gas surface densities in our simulations
at $250$ Myr for the different feedback models, and compare it with
the empirical relation shown as a solid line (normalised for a
Chabrier IMF, see \citealt{DallaVecchia2012}). In this plot we
include results from both the low-mass \sbc{} and high-mass \mw{}
galaxies, in order to show a wide range of surface densities, and to
demonstrate how the feedback efficiency changes for each model with
galaxy mass, metallicity, and physical resolution.  Results for the
\sbc{} galaxy are shown with smaller opaque symbols, while the \mw{}
galaxy is represented by larger and more transparent symbols. The gas
and SFR surface densities are averaged along annulli around the galaxy
centre, with equally spaced azimuthal bins of $\Delta r=500$ pc, and
we only include gas within a height of $2$ kpc from the disc plane
($4$ kpc in the case of the \mw{} disc).

All feedback models, and even the case of no feedback, produce slopes
in the KS relation in rough accordance with observations at gas
surface densities substantially above the `knee' at
$\Siggas \approx 10 \, \msunpc$, though the slopes tend to be slightly
steeper than observed. The similarity to the observed slope is in
large part a result of the built-in star formation model,
$\dot \rho_{*} \propto \rho^{1.5}$.  However, even though all
simulations have the same sub-resolution local star formation
efficiency of $\sfeff = 2\%$, the $\Sigsfr$ normalisation varies by
about an order of magnitude, with delayed cooling being most efficient
at suppressing the star formation for any given gas surface density,
owing to the large scale height of the disc. At high gas surface
densities ($\Siggas \gg 10 \, \msunpc$), all methods predict too high
$\Sigsfr$, except delayed cooling which predicts too low values.

For the lower-mass \sbc{} galaxy (smaller solid symbols), thermal
dump, stochastic, mechanical feedback, and delayed cooling are all
similar in the KS plot, though delayed cooling does not produce as
high gas surface densities as the other models. Kinetic feedback has
significantly higher SFR surface densities for given gas surface
densities (but relatively low maximum gas surface densities), owing to
the very thin disc produced by the almost decoupled injection of
momentum.

For the more massive \mw{} galaxy, which was simulated with lower
resolution, the picture is quite different (large transparent
symbols). With thermal dump feedback, the SFR surface densities shift
significantly upwards and the relation is quite similar to the no
feedback case. Stochastic feedback, and to a lesser extent, mechanical
feedback, also shift upwards, away from the observed relation. For
kinetic feedback, the relation is however almost unchanged in the more
massive galaxy (except for low gas surface densities, where it is
higher), but consistently remains about a factor two above the
observed relation. With delayed cooling, the gas surface densities
become much higher than in the lower mass galaxy, but the SFR surface
densities are significantly lower than observed.

For delayed cooling, we can calibrate the available free parameter to
improve the comparison to observations. Halving the delayed cooling
time-scale in the \mw{} galaxy, to the same value as used for the
\sbc{} galaxy, results in a KS relation which is very close to the
observed one. For the other models, we cannot calibrate the feedback
parameters to close in on the observed relation, and other measures
are required, such as increasing the feedback energy per unit stellar
mass. Another option is to reduce the star formation efficiency
parameter, $\sfeff$, in which case a fair match to observations can be
produced, but at the cost of making the feedback insignificant
compared to the no feedback case in terms of morphology, total SFR,
and outflows (the feedback essentially all becomes captured inside
$\sfeff$).

\subsection{Outflows} \label{outflows.sec} Galactic outflows are a
vital factor in delaying the conversion of gas into stars. Feedback
processes in the ISM are thought to eject large quantities of gas from
the galaxy, some of the gas escaping the gravitational pull of the
galactic halo altogether. Most of the gas, however, is expected to be
ejected at velocities below the halo escape velocity and to be
recycled into the disc. Galactic outflows are routinely detected in
observations \citep[e.g.][]{Steidel2010, Heckman2015}, and while
the outflow speed of cold material can be fairly accurately
determined, other properties of the outflows are not well constrained,
including the mass outflow rate, the fraction of gas escaping the
halo, the density, and thermal state of the gas.

Outflows are often characterised in terms of the mass loading factor,
which is the ratio of the outflow rate and the rate of star formation
in the galaxy. Its definition is somewhat ambiguous, as it depends on
the geometry and distance from the galaxy at which the outflows are
measured, which is hard to determine in observations.  Observational
works have inferred outflow mass loading factors well exceeding unity
\citep[see e.g.][]{Bland-Hawthorn2007a, Schroetter2015}, and many
theoretical models require mass loading factors of $1-10$ in sub-L$_*$
galaxies to reproduce observable quantities in the Universe
\citep[e.g.][]{Puchwein2013,Vogelsberger2013,Barai2015,Mitra2015}.

It is therefore important to consider outflow properties when
evaluating SN feedback models. Models that produce weak or no
outflows, with mass loading factors well below unity, could be at odds
with current mainstream theories of galaxy evolution (although it is
not known whether SN feedback is directly responsible for outflows --
e.g. cosmic rays could play a major role; \citealt{Booth2013,
  Hanasz2013, Salem2014, Girichidis2016}).

\begin{figure}
  \centering
  \includegraphics
    {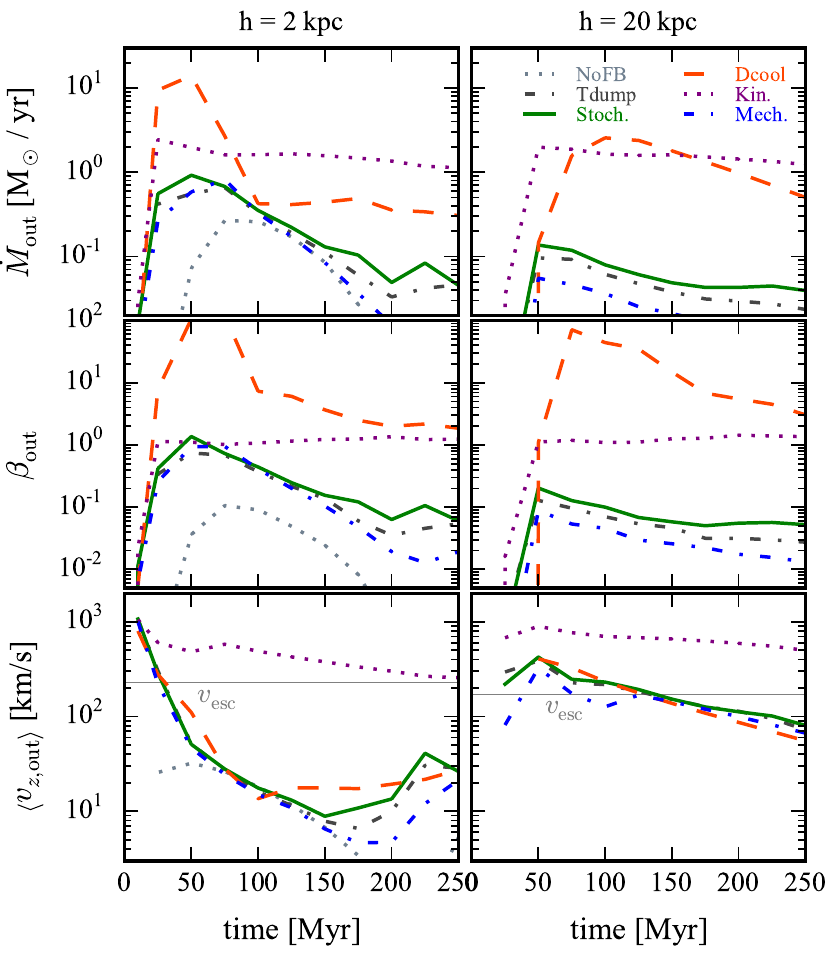}
  \caption
  {\label{OFtime.fig}Gross mass outflow rates ($\Mout$), mass loading
    factors ($\betaOut$), and mass-weighted average outflow velocities
    ($\vzout$), across planes $2$ and $20$ kpc above the disc plane of
    the \sbc{} galaxy (left and right columns, respectively), for the
    SN feedback models and their fiducial parameters. The colour
    coding and linestyles are the same as in \Fig{SFR.fig}. The thin
    horizontal lines in the bottom panels indicate the escape
    velocity.}
\end{figure}

In \Fig{OFtime.fig} we compare the time-evolution of outflows from the
\sbc{} galaxy with the different SN feedback schemes\footnote{We show
  outflow plots for the \sbc{} galaxy only, but we comment on outflows
  in the more massive \mw{} galaxy (which have similar properties) at
  the end of this subsection.}. We measure the gross gas outflow
(i.e. ignoring inflow) across planes parallel to the galaxy disc, at a
distance of $2$ kpc in the left panels and further out at $20$ kpc in
the right panels. The top row of panels shows the mass outflow rate
across those planes ($\Mout$), the middle row shows the mass loading
factor ($\betaOut$), and the bottom row shows the mass-weighted
average of the outflow velocity perpendicular to the outflow plane
($\vzout$).

In terms of outflows $2$ kpc above the disc plane (left panels of
\Fig{OFtime.fig}), kinetic feedback is strongest, with
$\Mout\approx 1 \, \msunyr$ and $\betaOut$ slightly above
unity. Delayed cooling produces a substantially lower outflow rate,
but since the SFR is also much lower, the mass loading factor is
higher. The other feedback models give much lower outflow rates, and
have mass loading factors $\sim 10^{-2}-10^{-1}$. At a larger distance
from the disc of $20$ kpc (right panels of \Fig{OFtime.fig}), the
situation is quite similar. All models except for kinetic feedback
have declining outflow rates and mass loading factors, owing to the
strong initial starburst, which can be seen to result in an outflow
rate peaking around $50$ Myr.

In the bottom panels of \Fig{OFtime.fig} we compare the average
outflow velocities to the DM halo escape velocity\footnote{The escape
  velocity estimate ignores the contribution of baryons. Hence, it is
  an underestimate, that is likely non-negligible close to the disc,
  but insignificant at $20$ kpc.},
\begin{align}
  \vesc(h) \approx 1.16 \ \vcirc \
  \sqrt{\frac{\ln \left(1+c x \right)}{x}},
\end{align}
where $x=h/\Rvir$ \citep[e.g.][]{Mo2004}, which has been marked
with horizontal grey solid lines.  Close to the disc, the average
velocity for kinetic feedback is marginally higher than escape, but
slowly declining due to the declining SFR. For the other feedback
models, the outflow velocity is well below escape. Ten times further
out, the mean outflow velocities are considerably higher for all
feedback models. This is to some degree due to the initial starburst,
which ejects high-velocity outflows early in the simulation runs, and
to some degree due to gas at lower velocities not having reached $20$
kpc and thus not contributing to the velocity average. In any case,
these (average) velocities are still below the escape velocity, again
with the exception of kinetic feedback. This implies that for all
models except kinetic feedback, most of the outflowing gas will not
escape to infinity, but instead fall back on the galaxy where it will
eventually produce stars. Kinetic feedback does give the gas high
enough velocity so that in principle it can escape the halo entirely,
while in practice this may be complicated by CGM and IGM gas which
stands in the way and needs to be swept out.

The main message of \Fig{OFtime.fig}, however, is not the escape
velocity, but the low mass loading factors for thermal dump,
stochastic, and mechanical feedback, far below the order unity
inferred from observations of local galaxies. As before, we see a
strong similarity between the results produced by thermal dump,
stochastic, and mechanical feedback.

\begin{figure}
  \includegraphics[width=0.49\textwidth]
    {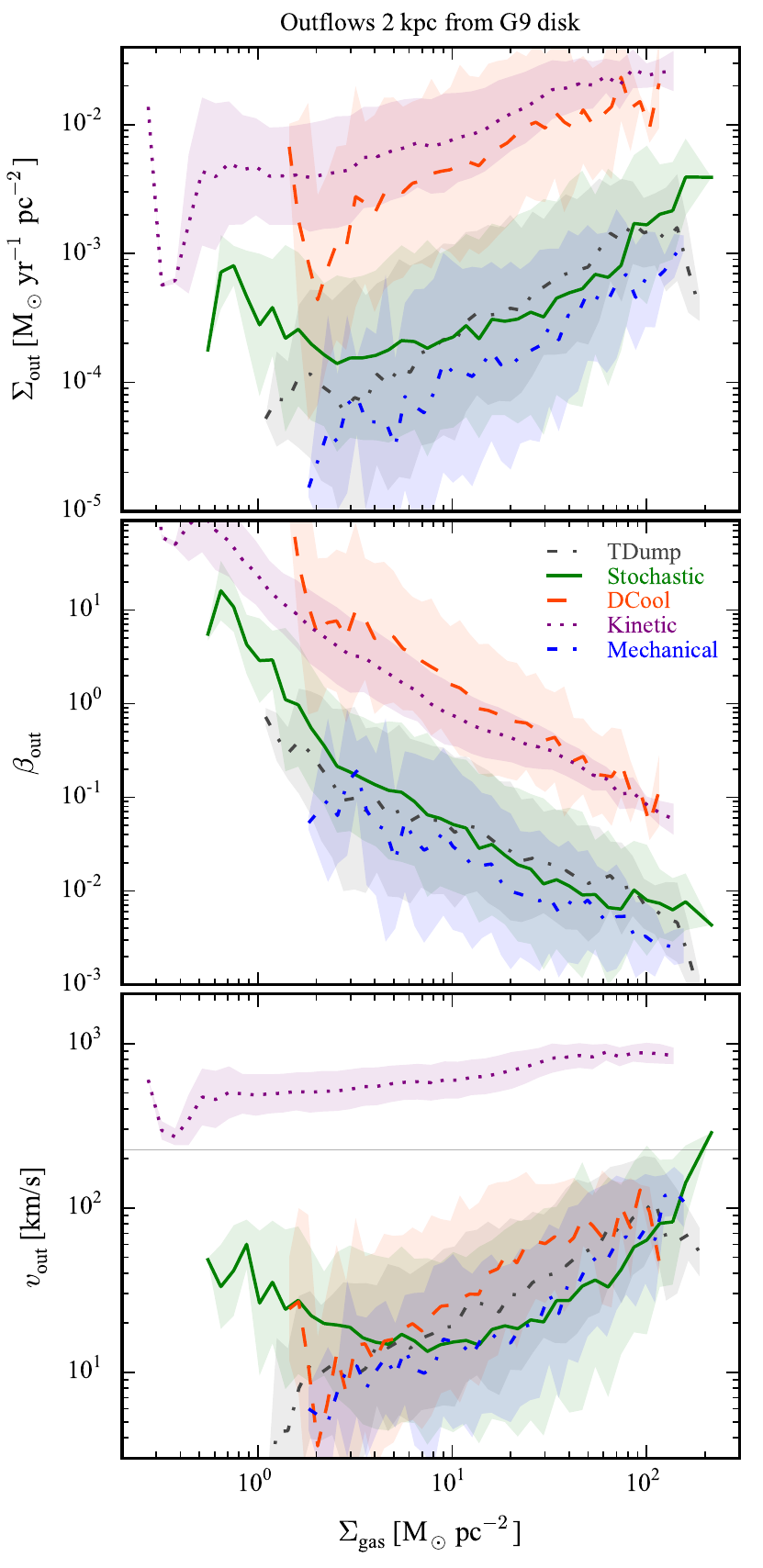}
  \caption
  {\label{SQO_Fid.fig}Local outflow properties at $250$ Myr across
    planes $2$ kpc from the \sbc{} disc as a function gas surface
    density, sampled from a $10$ kpc wide square grid of $100$ by
    $100$ squares in the xy-plane of the galaxy disc.  All curves are
    binned by gas surface density, with shaded regions showing
    standard deviations within each bin.  {\bf Top panel:} gross
    outflow rates per unit area. {\bf Middle panel:} mass loading
    factors, i.e. average mass outflow fluxes divided by star
    formation surface densities.  {\bf Bottom panel:} mass-weighted
    average gross outflow velocities (with the escape velocity shown
    by a horizontal solid line).}
\end{figure}

In \Fig{SQO_Fid.fig}, we study how the outflow properties $2$ kpc from
the disc scale with the local gas surface density. The panels show,
from top to bottom, gross outflow rate per unit area ($\SigOut$), mass
loading factor $\betaOut \equiv \SigOut/\Sigsfr$, and mass-weighted
gross outflow velocity.  We split the face of the disc into a $10$ kpc
wide grid of $100$ pc squares (that is, $100^2$ squares), and extract
the outflow properties in each square. In the plots, the outflow
properties are binned by the gas surface density, and the shaded
regions show the logarithmic standard deviation in each bin.

For each model, the general trend is that higher gas surface
densities correspond to higher outflow rates and velocities, but lower
mass loading factors.  The outflow velocities are noticeably higher
than those (at $250$ Myr) in \Fig{OFtime.fig}. The reason is that
\Fig{OFtime.fig} shows the mass-weighted average of all outflowing
gas, while \Fig{SQO_Fid.fig} is restricted to a $10$ kpc wide square
plane directly above and below the disc, hence capturing the more
collimated part of the outflows.  Kinetic feedback clearly stands out
as having the highest outflow velocities, peaking close to $10^3$ km/s
(at the peak surface densities), with little scatter. With kinetic
feedback, almost all of the outflowing gas directly above or below the
disc is moving faster than the escape velocity (indicated with a
horizontal solid line). The other feedback models produce outflow
rates and velocities that are alike (within roughly a factor of two),
and much lower than for kinetic feedback, with the exception of
delayed cooling, which has a massive, slow outflow, and the highest
mass loading factor.

\begin{figure*}
  \centering
  \subfloat
  {\includegraphics[width=0.3\textwidth]
    {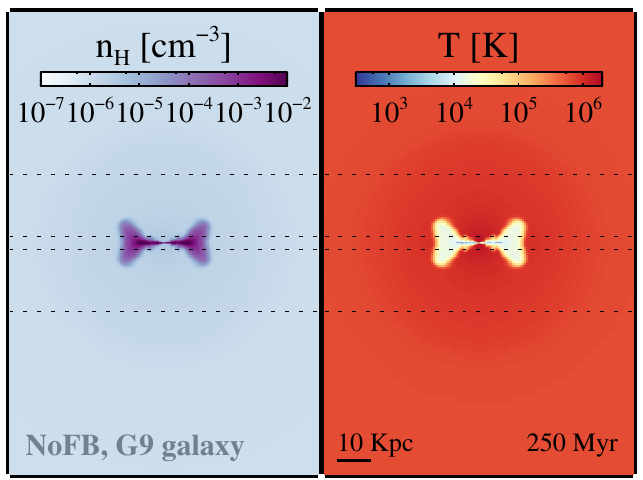}}
  \subfloat
  {\includegraphics[width=0.3\textwidth]
    {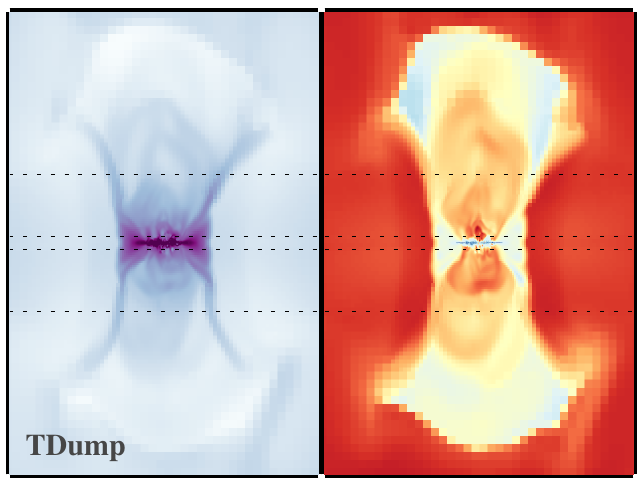}}
  \subfloat
  {\includegraphics[width=0.3\textwidth]
    {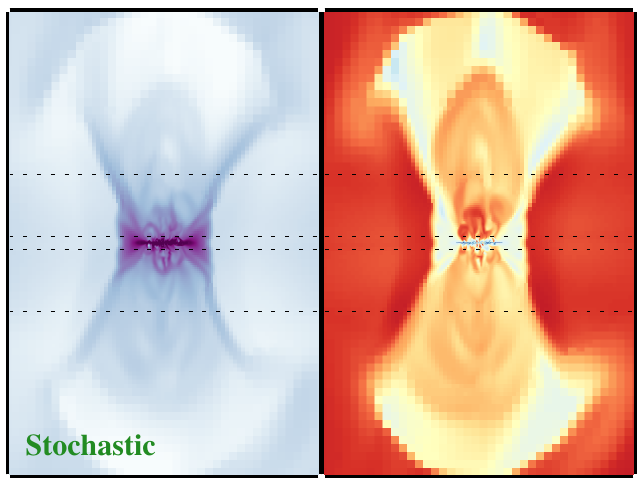}} \\
  \vspace{-4mm}
  \subfloat
  {\includegraphics[width=0.3\textwidth]
    {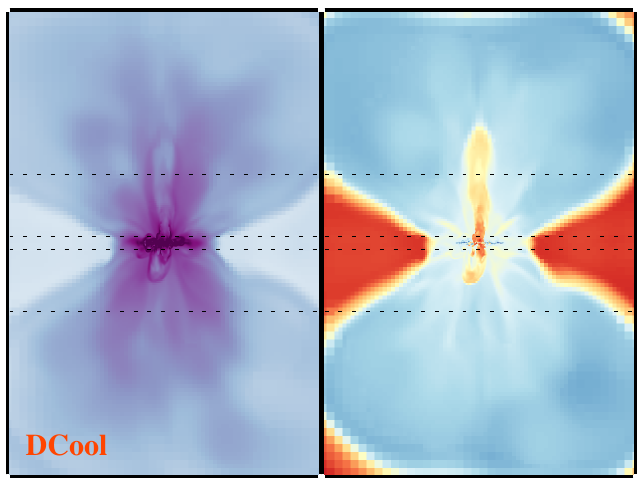}}
  \subfloat
  {\includegraphics[width=0.3\textwidth]
    {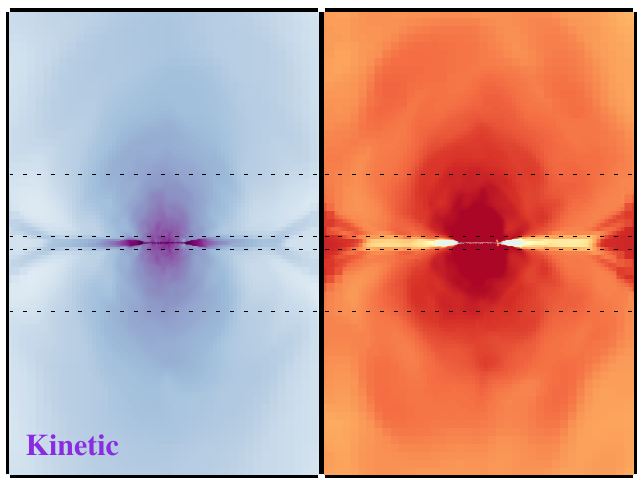}}
  \subfloat
  {\includegraphics[width=0.3\textwidth]
    {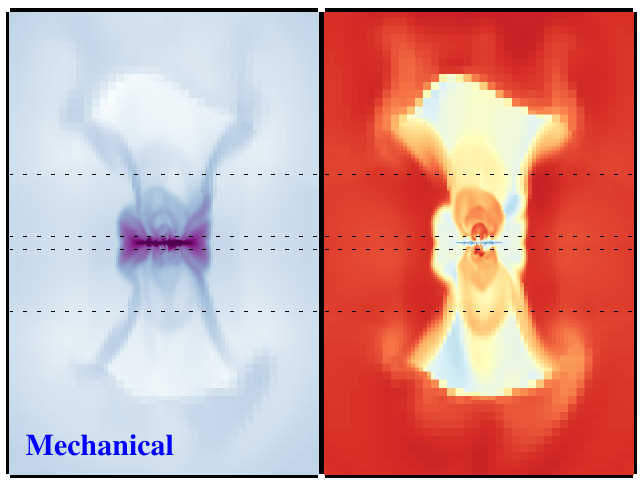}}
  \caption
  {\label{mapsOF_fid.fig}Slices along the $xz$-axes (at $y=0$; the
    disc is seen edge-on), for the \sbc{} galaxy at $250$ Myr. Each
    panel of two images shows the hydrogen number density (left) and
    the gas temperature (right) for a given model. The models are,
    clockwise from the top left (as indicated in the lower left corner
    of each density map): no feedback, thermal dump, stochastic,
    mechanical, kinetic, delayed cooling. In each map, dotted
    horizontal lines mark planes at which we measure the outflow
    properties shown in Figures \ref{OFtime.fig} and
    \ref{SQO_Fid.fig}, i.e. at $2$ and $20$ kpc from the plane of the
    disc. Thermal dump, stochastic, and mechanical feedback produce
    qualitatively similar multi-phase outflows. Delayed cooling
    produces outflows that are dense, cold, and slow, whereas those
    produced by kinetic feedback are diffuse, hot, and fast.}
\end{figure*}

In \Fig{mapsOF_fid.fig} we show images of slices along the $xz$-plane
(at $y=0$) at $250$ Myr, with each set of two panels showing the
hydrogen density (left) and temperature (right) for a given feedback
model. Delayed cooling and kinetic feedback clearly stand out here.
The former yields dense and cold ($T\la10^4$ K) outflows. The outflows
for the latter are diffuse, hot
($10^6 \ {\rm K} \, \la T \la 10^8 \ {\rm K}$), and the most extended
(which is expected since \Fig{SQO_Fid.fig} shows they are by far the
fastest). The remaining three feedback models produce qualitatively
similar multiphase ($10^4 \ {\rm K} \, \la T \la 10^6$ K)
outflows. The clear distinction between the most effective feedback
model, i.e. delayed cooling, and the other, less effective models,
in the outflow properties, could be used in future work as an
observational probe into how accurately those models represent actual
feedback in galaxies.

\subsubsection{Outflows from the more massive galaxy}
For the more massive \mw{} galaxy, the outflow differences, which we
do not plot, are qualitatively similar to those in the above
analysis. Kinetic feedback gives the highest mass loading factor,
which is again of order unity both at $2$ and $20$ kpc. All the other
models give similar mass loading factors $2$ kpc above the disc as
for \sbc{}, but in contrast to \sbc{} the mass loading drops by $1-2$
orders of magnitude at $20$ kpc (the biggest drop occurring for
delayed cooling), due to the stronger gravitational pull. The outflow
velocities are slightly higher for all models, but they are still
much (marginally for kinetic feedback) lower than the ($\approx 500$
km/s) escape velocity.

\subsection{Gas properties}
\begin{figure*}
  \centering
  \includegraphics[width=0.9\textwidth]
    {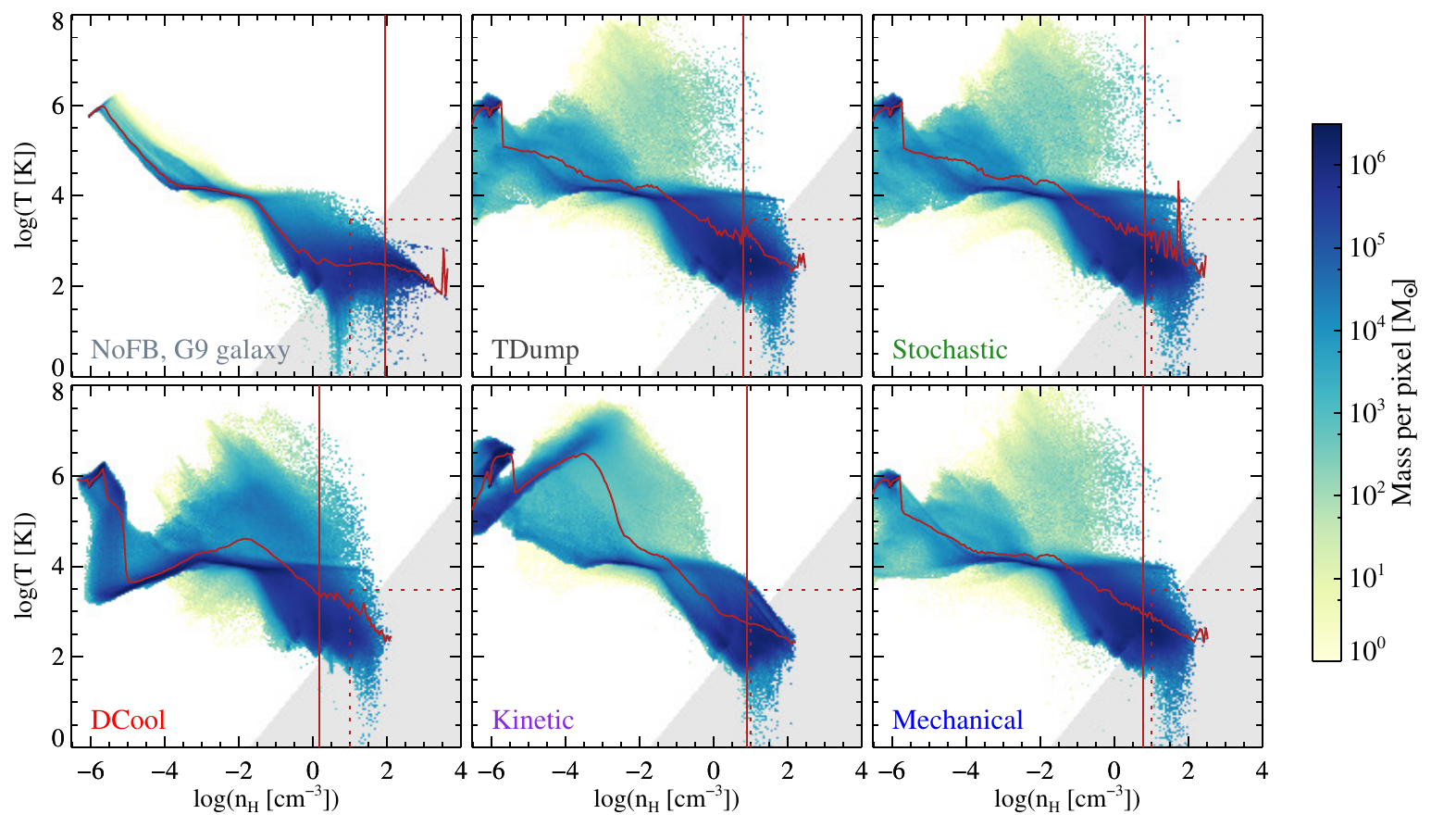}
  \caption
  {\label{PH_Fid.fig}Phase diagrams at $250$ Myr for \sbc{} runs with
    various feedback models and fiducial settings. The shaded grey
    region marks where the temperature is below the Jeans temperature
    which is added artificially for pressure support. Dashed
    red horizontal and vertical red lines enclose gas that is
    star-forming. The solid vertical red line in each plot marks the
    (mass-weighted) mean density, and the red solid curve shows the
    mean temperature as a function of density. The diagrams are almost
    identical for thermal dump, stochastic, and mechanical
    feedback. For delayed cooling, we find a lot of dense gas at
    `forbidden' temperatures ($\sim 10^5$ K), where the cooling rate
    peaks.}
\end{figure*}
In \Fig{PH_Fid.fig} we compare gas properties for runs with the
different SN feedback models, using phase diagrams of gas temperature
versus density at $250$ Myr. The colour scheme represents the mass of
gas in each temperature-density bin. The mass-weighted mean density in
each diagram is represented by a solid vertical line, while the
mass-weighted mean temperature in each density bin is shown by solid
red curves. Star-forming gas is enclosed by a dotted box, while gas
with temperatures below the artificial Jeans temperature (which has
been subtracted from the `thermal' temperature plotted here) is
indicated by the shaded diagonal region in the bottom right corner of
each diagram.

We continue to see the same qualitative picture as before: delayed
cooling and kinetic feedback stand out, while the remaining three
models look similar. Delayed cooling yields by far the lowest mean
density, $\left< \nh \right> \approx 3 \, \cci$, which is well below
the star formation density threshold of $\ns=10 \ \cci$, and almost
two orders of magnitude below the mean density without feedback. The
other models all yield mean densities near $\ns$.

Delayed cooling produces the highest mean temperatures at intermediate
densities of $\nh = 10^{-2}-10 \ \cci$ with a lot of gas at
temperatures $10^{4.5}-10^6$ K, which is probably unphysical given the
short radiative cooling times in this regime. Curiously, in more
diffuse gas, $\nh = 10^{-5}-10^{-3} \, \cci$ delayed cooling produces
by far the lowest temperatures, $T\sim 10^3-10^4$ K, while in the same
density regime the gas is typically at $\sim 10^5$ K with other
feedback models (even in the absence of feedback). From identical
phase diagrams excluding the galaxy disc, we have confirmed that this
diffuse gas is primarily `CGM' gas outside the disc. In the case of
delayed cooling, these outflows, i.e. gas outside the disc, span a
wide range of densities,
$\nh \sim 10^{-6} - 3 \times 10^{-1} \, \cci$, in stark contrast to
the other feedback models, where the CGM gas reaches maximum
densities of a few times $10^{-3} \, \cci$.  With kinetic feedback,
the CGM has a clear bi-modality not seen for other models, with some
of the gas following an adiabat starting around
$\nh\sim 10^{-2.5} \, \cci$, $T\sim 10^{7.5}$ K, and extending towards
much lower densities, and the remainder at $T \ga 10^4$ K and spanning
densities of $\nh \sim 10^{-5} - 3 \times 10^{-3} \, \cci$ (the latter
is flowing 'diagonally' from the disc, i.e. at a steep angle from the
axis of disc rotation).

All feedback models produce hot and relatively diffuse gas,
populating the region $T\sim10^4-10^8$ K,
$\nh\sim 3 \times 10^{-4} - 10 \, \cci$ in \Fig{PH_Fid.fig}. One might
expect this to belong to the outflowing CGM. However, if we exclude
the disc (out to $2$ kpc in height and $10$ kpc in radius), all of
this hot diffuse gas disappears from the phase diagrams, indicating
that it in fact belongs to the ISM. In the case of thermal dump,
stochastic, and mechanical feedback, the outflowing CGM is indeed warm
to hot, but dilute, with $\nh\sim 10^{-6}-10^{-4} \, \cci$, while
kinetic feedback produces the aforementioned bi-modality in the
outflowing CGM, and delayed cooling produces circum-galactic outflows
that are predominantly at temperatures between $10^4$ and $10^5$ K.

\subsection{Impact of feedback on the local
  environment} \label{SD.sec} A major factor in any feedback model is
how efficiently it clears away those dense clouds where stars can
form. When dense regions are quickly cleared by early SN explosions in
a stellar cluster, this can also boost the efficiency of subsequent SN
explosions which then take place at lower densities where cooling is
less efficient and the momentum obtained in the SN remnant increases
\citep[Eq. \ref{dp_mech.eq}; see also][]{Kim2014, Martizzi2015}.
SNe exploding in the diffuse ISM have been suggested to prevent the
formation of star-forming clouds \citep[e.g.][]{Iffrig2015}, to
maintain the hot volume-filling ISM, and to generate fast outflows
\citep[e.g.][]{Ceverino2009}.

\begin{figure}
  \centering
  \includegraphics
    {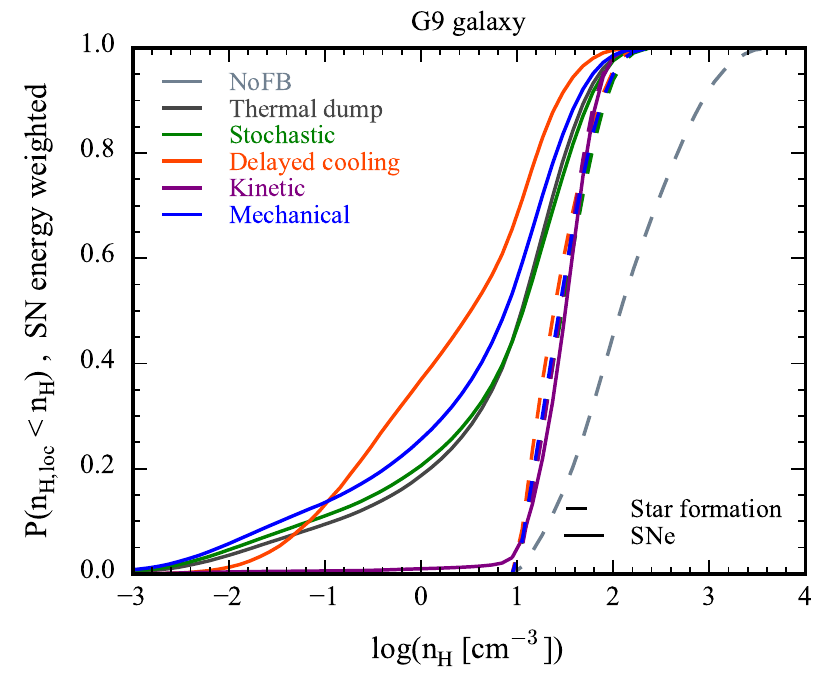}
  \caption
  {\label{SD_cum.fig}Local densities at which stellar particles are
    formed over the run-time of the \sbc{} galaxy (dashed curves) and
    at which subsequent SN events take place 5 Myr later (solid
    curves), colour coded by feedback model as indicated in the upper
    legend. For all of the feedback models studied, star formation
    takes place at densities close to the density threshold for star
    formation, $\ns=10 \ \cci$, which is significantly lower than the
    densities at which stars form without feedback. Except for
    kinetic feedback, most SN events happen at much lower densities,
    indicating that the local environment is significantly altered by
    feedback.}
\end{figure}

In \Fig{SD_cum.fig} we show the gas densities at which stellar
particles are born (dashed curves) and the densities at which the SN
events take place (solid curves) for each of the feedback models in
the \sbc{} galaxy. Focusing first on the star formation densities,
they are almost indistinguishable for all the models, and differ only
for the case of no feedback, for which the stars typically form at
significantly higher densities. This shows, as we have already seen
from the morphological comparison in \Fig{maps_Fid.fig}, that all the
feedback models are efficient at preventing and/or destroying
star-forming clouds in this \sbc{} galaxy, and almost all the stellar
particles are formed within one dex of the star formation density
threshold of $\ns=10 \ \cci$. For the no feedback case, the clouds can
collapse to higher densities, impeded only by the density-dependent
pressure floor.

For the densities at which the SN events take place, there are larger
differences between the feedback models. Thermal dump, stochastic,
and mechanical feedback are similar, with a non-negligible
$\approx 10$ per cent of the SN energy injected below $0.1 \ \cci$
($20-40$ per cent below $1 \ \cci$), and SN events consistently taking
place at lower densities than star formation.

Delayed cooling stands out, in having more SNe than those
aforementioned models at densities $\ga 10^{-1} \ \cci$, but fewer
SNe for lower densities. This comes from the efficiency of delayed
cooling in diffusing and thickening the ISM disc, resulting in a
mass-weighted density distribution of the ISM (not shown) which peaks
at $\nh \approx 1 \ \cci$, about a dex lower than for thermal dump,
stochastic, and mechanical feedback, but a volume-weighted
distribution (also not shown) which peaks at
$\nh \approx 10^{-2} \ \cci$, about a dex \emph{higher} than for those
other models. Delayed cooling hence smooths out not just the density
peaks in the ISM, but also the density throughs, such that stars form
at lower average densities, but SNe explode at higher minimum
densities than for the other models.

Standing out much more distinctly, kinetic feedback SNe explode in gas
with almost exactly the same densities (and even higher) as the stars
are formed, with almost no SNe exploding at lower densities. This
signature of kinetic feedback suggests a decoupling between the
injected momentum and the immediate environment surrounding the
exploding stellar particle. Instead of quickly dispersing the sites of
star formation, the gas is gradually transported away from those
sites, out of the ISM, without coupling to the immediately surrounding
gas. This explains the thin bubble-free gas disc (\Fig{maps_Fid.fig})
and the gradual decline of the SFR (\Fig{SFR.fig}), which is due to
slow gas depletion. These distinct features are however strongly
dependent on the size of the bubble, $\rb$, into which the SN momentum
and mass is injected. The fiducial size $\rb=150$ pc results in this
decoupling between the SNe and the surrounding gas. In
\Sec{fb_kin.sec} we experiment with a smaller bubble size ($\rb=40$
pc) and find a very different behaviour for kinetic feedback, with
results resembling those for thermal dump, stochastic, and mechanical
SNe, i.e. much lower outflow rates, a flatter star formation rate with
time, and a thicker disc.

\subsection{Similarity of three models in the low-mass
  galaxy} \label{resolved.sec}

For the low-mass \sbc{} galaxy (but not for the more massive \mw{}
galaxy), we have seen that the results for thermal dump, stochastic,
and mechanical feedback are near identical in terms of morphologies,
star formation, and gas properties.

In \Eq{pSN.eq} we showed that the probability for stochastic feedback
events is above unity at the density threshold for star formation
($\ns=10 \, \cci$) in the \sbc{} galaxy, and in
\Eq{chi_mech_approx.eq} we saw that, also at $\ns$, mechanical
feedback remains in the resolved adiabatic phase. In addition, we
found in \Sec{SD.sec} that SN events do typically take place at
densities close to $\ns$. Hence there appears to be no significant
numerical overcooling issue in the \sbc{} runs, and it is then no
surprise to find similar results for thermal dump, stochastic, and
mechanical feedback. It can be argued that the adiabatic phase of
thermal dump feedback is resolved. Note that this may imply that
delayed cooling and kinetic feedback, for the fiducial parameters we
have chosen, converge to wrong results for the effects of SN feedback.

For the more massive \mw{} galaxy, this is not the case: these
aforementioned feedback models give quite different results
(\Fig{SFR_MW.fig}), and thermal dump does not do much to suppress star
formation compared to the no feedback case. Purely from Equations
\ref{pSN.eq} (stochastic probability) and \ref{chi_mech_approx.eq}
(mechanical feedback phase), one might expect the adiabatic phase to
be resolved here as well, since the changes in stellar mass and
resolution, compared to the \sbc{} galaxy, cancel out. However, in
part due to stronger metal cooling, but more importantly due to the
larger gravitational potential of the \mw{} galaxy, stars form and
explode at significantly higher densities than in the \sbc{}
galaxy. Hence the probability for stochastic feedback events becomes
lower than unity (on average $0.35$ in the stochastic feedback run),
most mechanical feedback events become pure snowplow momentum
injections, and thermal dump becomes a victim of numerical
overcooling.

\cite{Kim2014} derived a density limit at which the momentum
created by single thermal dump type II SN explosions is
resolution-converged with grid-hydrodynamics. They found that
convergence is maintained with a cell width
$\Delta x \la 10 \, {\rm pc} \ n_0^{-0.46}$, where
$n_0=\frac{\nh}{1 \, \cci}$. Taking $\ns=10 \, \cci$ we would need a
resolution of $\approx 3.5$ pc for converged thermal dump feedback.
That is a considerably higher resolution than ours ($18$ pc), so
naively we would expect overcooling to be significant for the \sbc{}
thermal dump simulation.  In light of the above finding, that thermal
dump feedback appears more or less resolved in the \sbc{} galaxy, the
lack of resolution according to \cite{Kim2014} is mitigated by the
fact that each stellar particle in our \sbc{} simulations releases the
equivalent of 40 type II SN explosions instantaneously, instead of
one, as assumed in \cite{Kim2014}.

\subsection{SN model comparison summary}

\begin{itemize}
\item For the low-mass \sbc{} galaxy, the results for thermal dump,
  stochastic, and mechanical feedback are near identical in terms of
  morphologies, star formation, and gas properties. This is an
  indication that the adiabatic phase of SN explosions is
  resolved. For the more massive \mw{} galaxy, thermal dump is
  significantly weaker than stochastic and mechanical feedback in
  suppressing star formation (though not so much in generating
  outflows).
\item Delayed cooling is by far most efficient at suppressing star
  formation and yields results closest to the observed
  Kennicutt-Schmidt relation (at least for our assumed star formation
  efficiency.
\item Thanks to a large fiducial `bubble radius' of $150$ pc for
  momentum and mass injection, kinetic feedback has the highest
  outflow rates and a mass loading factor of order unity. Delayed
  cooling follows, with weaker outflow rates but a slightly higher
  (but declining) mass loading factor. The other feedback models
  produce much lower outflow rates and mass loading factors than those
  two more efficient models. In the more massive ($\sim$ MW) \mw{}
  galaxy, the mass loading factor is similar to that of \sbc{} close
  to the galaxy plane, but drops by $1-2$ orders of magnitude ten
  times further out at $20$ kpc, for all models except kinetic
  feedback, which maintains a mass loading factor of unity.
\item The feedback models producing the lowest outflow rates and mass
  loading factors produce hot and dilute outflows, while delayed
  cooling yields distinctively cold and dense outflows. For kinetic
  feedback the outflows have a clear bi-modal phase structure, with
  relatively cold and dense outflows close to the disc, and hot and
  diffuse outflows further out following a temperature-density
  relation suggesting adiabatic cooling.
\item Given the large fiducial bubble radius, which effectively
  decouples feedback from the ISM, kinetic feedback produces by far
  the fastest outflow, some of it above the escape velocity. All other
  models produce outflow velocities about an order of magnitude
  lower, well below the escape velocity.
\end{itemize}

\section{Resolution convergence} \label{res_conv.sec} Resolution
convergence is an important factor in assessing SN feedback
models. Ideally, the effects of feedback should remain constant if the
resolution is increased, or at least if it is varied within reasonable
limits, i.e. within a factor of a few\footnote{Convergence with a
  dramatic change in resolution, on the other hand, is usually not a
  desired goal. With much lower resolution, a sub-grid model becomes
  meaningless as the structures with which the model is supposed to
  interact become completely unresolved and a `lower level' of
  sub-grid physics must take over the current ones. With much higher
  resolution, some of the real physics become resolved, and the
  sub-grid model becomes irrelevant (though ideally it should then
  converge towards a `first-principles' methodology).}. In practice
such constancy, while desirable, is hard to obtain without making
significant sacrifices, such as disabling physical processes like
hydrodynamical interactions or radiative cooling in the ISM. A second
best choice is a small and predictable change with varying resolution,
so the feedback parameters can be easily calibrated for different
setups in order to achieve ``weak convergence'' \citep{Schaye2015}.

In this section we aim to understand how and to what extent measurable
galaxy properties change with resolution for the different feedback
models. For this purpose, we use a lower-resolution version of the
\sbc{} galaxy, which we call \sbclr{}. The setup is identical to
\sbc{}, except that the minimum cell width is two times larger,
i.e. $36$ pc, and the particle mass (in the initial conditions as well
as for new stellar particles) is eight times higher (i.e.
$1.6 \times 10^4 \ \Msun$ for stellar particles,
$\approx 8 \times 10^5 \ \Msun$ for DM particles). For simplicity, and
because the Jeans length is already resolved by $7$ cell widths in the
higher-resolution runs (and hence by $3.5$ cell widths in the
lower-resolution ones), we leave the pressure floor and star formation
threshold unchanged.

\begin{figure}
  \centering
  \includegraphics
    {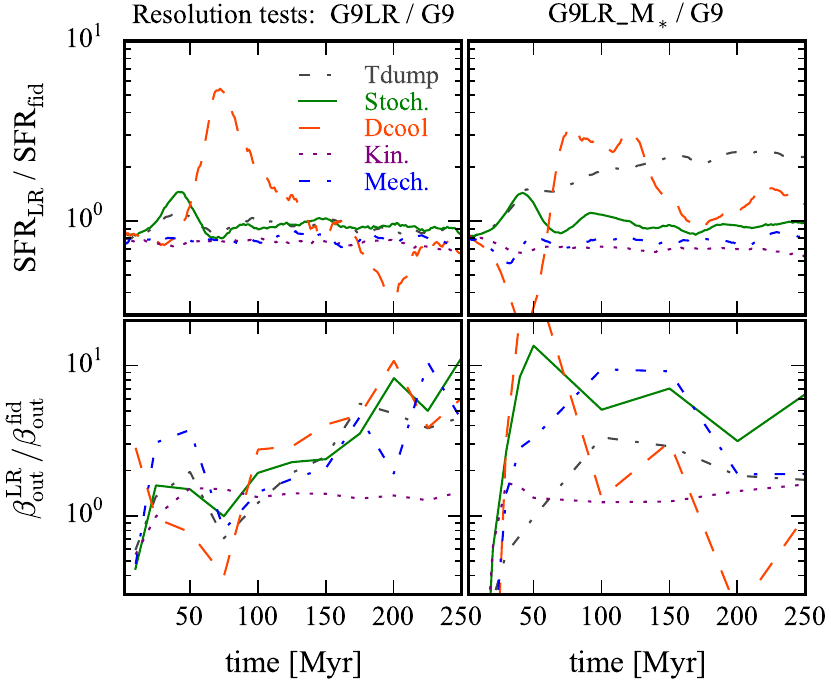}
  \caption
  {\label{SFR_res.fig}Resolution convergence tests. The upper (lower)
    panels show, for the different feedback models, ratios of SFRs
    (mass loading factors at 2 kpc from the disc) of \sbc{} runs at
    low resolution and at the fiducial resolution. The left (right)
    panels shows ratios where the stellar particle mass is increased
    by a factor of eight (kept fixed) in the low-resolution runs. For
    all plots, we have averaged the SFRs over intervals of $20$
    Myr. For low resolution with more stellar massive particles, the
    SFRs are well converged except for delayed cooling, though with a
    trend of marginally lower SFRs. The mass loading factors do change
    (except for kinetic feedback), generally showing an increase with
    lower resolution. With a fixed stellar particle mass (i.e. lowered
    SN specific energy), thermal dump feedback becomes weaker with
    lower resolution, while other feedback models maintain similar
    SFRs but higher mass loading factors.}
\end{figure}

In the left column of panels in \Fig{SFR_res.fig}, we plot, for each
feedback model, the ratios of SFRs (upper panels) and mass loading
factors $2$ kpc from the disc (lower panels) for \sbclr{} over \sbc{}
runs.

For delayed cooling, the resolution has a significant effect on the
relative SFR, although it should be kept in mind that the SFR for
delayed cooling is quite small in the first place (and hence the
absolute change is low compared to the SFRs of other models). For
other models, the SFRs change insignificantly with resolution, though
there is a systematic tendency of slightly lower SFR with lower
resolution.

The lower resolution has a larger effect on the outflow rates, shown
in the bottom left panel of \Fig{SFR_res.fig}. Decreasing the
resolution systematically increases the outflow rates for mechanical
feedback (by up to a factor four), thermal dump, stochastic feedback,
and delayed cooling, (by roughly a factor two), while for kinetic
feedback the outflow rate is almost unchanged. The outflow rate
increase is likely connected to the winds being launched on larger
scales, due to the larger cell width and mass (note that the specific
energy, i.e. the ratio between the SN energy and receiving gas mass is
similar to the higher resolution run, since the particle mass is $8$
times larger).  The obvious exception is kinetic feedback, where the
energy is distributed within a bubble radius which we have kept
constant, and indeed the outflow rate remains unchanged.

We also consider the effect of `SN' resolution, where we lower the
grid resolution (and that of the initial conditions particles), just
as in the \sbclr{} runs, but keep the mass of newly formed stellar
particles fixed compared to the \sbc{} runs. These runs, which we call
\sbclrms{}, give another measure of resolution convergence for the
feedback models, as the specific energy per feedback event (i.e. SN
energy over the local gas mass) is reduced by a factor of eight
compared to \sbc{} runs, while it was kept constant in
\Fig{SFR_res.fig}.  We compare those to the \sbc{} runs in the right
column of panels in \Fig{SFR_res.fig}, showing the ratios of SFRs and
mass loading factors for the feedback models.  For the SFR, the
largest difference occurs for thermal dump feedback, which becomes
much less efficient at suppressing star formation rates due to the
lower particle masses. The adiabatic phase becomes severely
unresolved, and there is no mechanism built into the model to
compensate.  Other feedback models maintain similar average star
formation with the lower specific energies.  The mass loading factors
increase somewhat if we decrease the spatial resolution and the
specific SN energies (bottom right panel of \Fig{SFR_res.fig}), but at
$250$ Myr the increase is smaller than for fixed specific energies
(bottom left panel). Delayed cooling is an exception, showing a
decrease in mass loading at some time intervals, but an increase in
others, which is likely just caused by the relatively large variations
in the SFRs.

With thermal dump feedback, resolution non-convergence is a well known
problem. With lower resolution, a larger gas mass is heated in a
single feedback event, resulting in lower initial temperatures given a
fixed SN energy. This would normally lead to higher SFRs, but in the
left panels of \Fig{SFR_res.fig} the effect is counter-balanced by the
use of more massive stellar particles, slightly increasing the
feedback efficiency due to the higher SN energy per feedback event.

For the case of stochastic feedback, the fairly good convergence of
the SFR with both spatial resolution and stellar particle mass is not
a big surprise, since the stochasticity is built in to ensure that the
heated gas receives a fixed specific energy, regardless of cell size,
density, and particle mass. The outflow rates are relatively poorly
converged for stochastic feedback and low stellar particle mass, which
likely comes from the aforementioned tendency for the outflow rate to
increase with lower resolution, and hence larger launching scales,
even if the specific SN energy is constant.  Mechanical feedback was
shown by \cite{Kimm2014} to converge well with resolution in terms
of the final momentum reached, and indeed the whole point of the model
is to maintain the same momentum injection regardless of whether the
momentum buildup is captured numerically or not. While we confirm that
the convergence is good for the SFR, the outflow rates are not very
well converged, neither in terms of spatial resolution nor stellar
particle mass. For delayed cooling, resolution convergence is not an
obvious property, but if cooling is turned off long enough, the energy
(i.e. cooling) losses should become insignificant and hence
independent of the resolution. The fact that the SFR (and to a lesser
extent the mass loading factor) is poorly converged for delayed
cooling hints that cooling losses are still significant, but we remind
that the absolute change is actually lower than for the other feedback
models.  As long as kinetic feedback is sufficiently decoupled from
the ISM surrounding the SN explosion due to the use of a large SN
bubble radius, it is not surprising to see good resolution
convergence, since the main effect of the feedback is then simply to
slowly deplete the disc of gas mass.

In summary, except for thermal dump feedback, the feedback models
converge fairly well in terms of SFRs. However, none except for
kinetic feedback (with a bubble radius larger than the disc height)
converge well in terms of the outflow mass loading factor, with the
mass loading factor decreasing with higher resolution.

\section{Varying the temperature jump for stochastic feedback} \label{fb_stoch.sec}

We now examine the effect of varying the stochastic heating,
parameter, $\DTstoch$. We do not go lower than the fiducial value of
$\DTstoch=10^{7.5}$ K, because already here the feedback is not very
stochastic: in the \sbc{} galaxy the average probability for SN
candidates (which, as a reminder, is the ratio of the available SN
energy of the stellar particle to the energy required to heat the host
gas cell by $\DTstoch$) is $\pSN\approx 50\%$ ($\approx 35\%$ in
\mw{}). Lower $\DTstoch$ would lead to order unity probabilities for
SN explosions, converging towards thermal dump feedback.

We thus consider three values for $\DTstoch$ in addition to the
fiducial one, each time increasing the injected specific energy by
half a dex, i.e. $\DTstoch=10^{7.5}, 10^{8}, 10^{8.5}$, and $10^{9}$
K.

\begin{figure}
  \centering
  \subfloat
  {\includegraphics[width=0.23\textwidth]
    {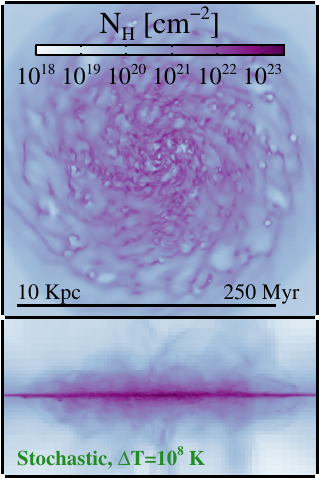}}
  \subfloat
  {\includegraphics[width=0.23\textwidth]
    {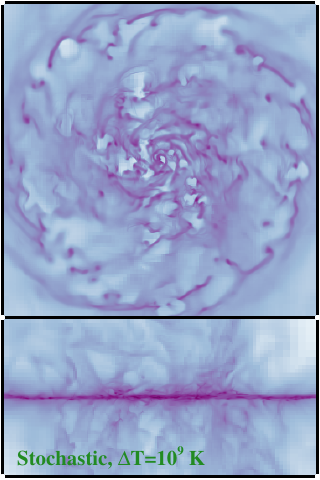}}
  \caption
  {\label{maps_stoch.fig}Maps of total hydrogen column density for the
    \sbc{} galaxy at $250$ Myr, for variations in the stochastic
    heating, with $\DTstoch=10^8$ K on the left and $\DTstoch=10^9$ K
    on the right. Each panel shows face-on and edge-on views.}
\end{figure}
\Fig{maps_stoch.fig} shows maps of the hydrogen column density for
$\DTstoch=10^8$ and $10^9$ K. For the case of $\DTstoch=10^8$ K, there
is not a significant difference from the fiducial run (the middle left
panel in \Fig{maps_Fid.fig}), though the gas disc becomes slightly
thicker and clumpier. For $\DTstoch=10^9$ K, the difference is much
clearer. Here the face-on disc is more diffuse overall, but at the
same time it contains more dense clumps and filaments, especially
close to the centre and at the disc outskirts. This is due to the
increased stochasticity of SN explosions: the low probability for SN
events, on average $\pSN\approx 3 \%$, allows dense star-forming
clumps to live longer before they are hit by the first SN
explosion\footnote{We note that \cite{DallaVecchia2012} advise
  against such high values of $\DTstoch$ that probabilities for
  feedback events become $\ll 1$, which is clearly the case
  here.}. This effect is amplified at the outskirts, where there is
relatively little star formation and thus a low rate of SN explosions
per unit volume. Looking at the disc edge-on, there is much more
structure in the CGM for our maximum value of $\DTstoch$.

\begin{figure}
  \centering
  {\includegraphics[width=0.45\textwidth]
    {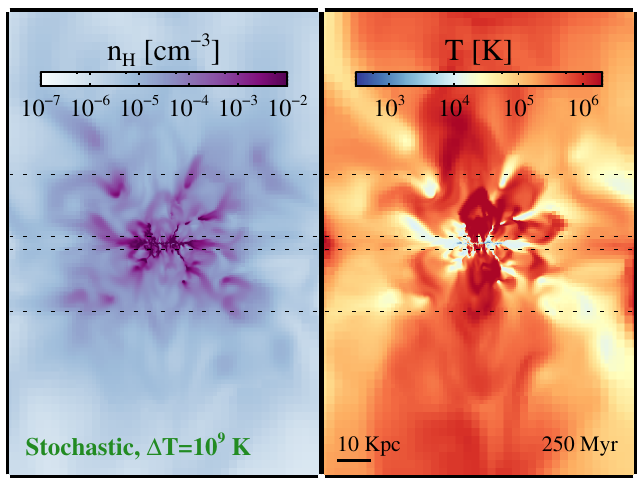}}
  \caption
  {\label{OF_stoch.fig}Edge-on views of the \sbc{} galaxy at $250$
    Myr, for the run with the strongest stochastic heating
    ($\DTstoch=10^9$ K). The two maps show slices of the hydrogen
    number density (left) and temperature (right). Dotted horizontal
    lines mark planes $2$ and $20$ kpc from the disc.}
\end{figure}

\begin{figure}
  \centering
  \includegraphics
    {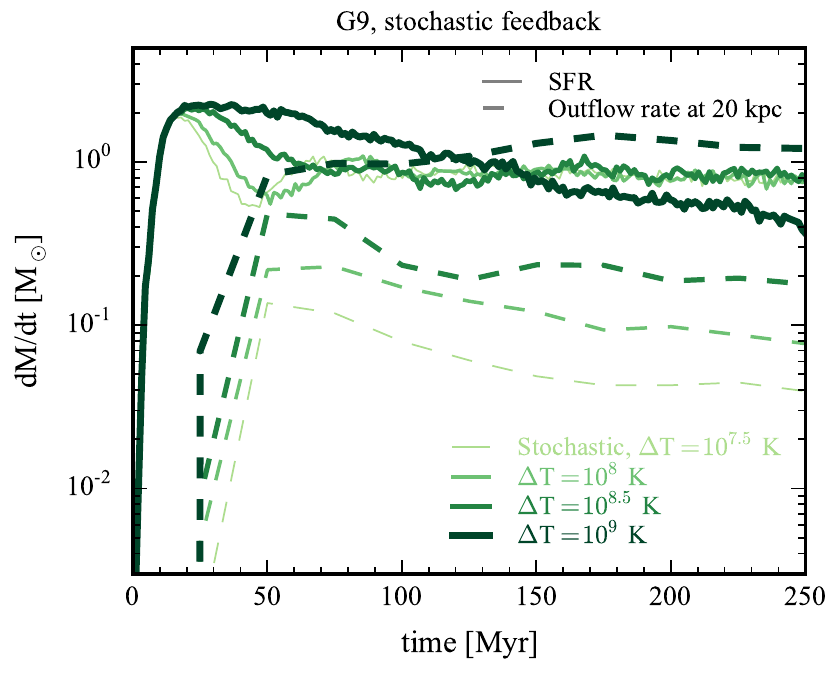}
  \caption
  {\label{SFR_stoch.fig}SFRs (solid curves) and gross outflow rates
    $20$ kpc from the disc (dashed curves) for the \sbc{} galaxy, with
    variations in $\DTstoch$ for the stochastic SN feedback model.
    Increasing $\DTstoch$ within reasonable limits has little effect
    on the SFR, but does increase the outflow rate (and hence the mass
    loading factor) significantly.}
\end{figure}

In \Fig{OF_stoch.fig} we zoom out and look at the large-scale outflows
in the `maximum' case of $\DTstoch=10^9$ K. The outflows are
dramatically different from the fiducial setting for stochastic
feedback shown in the top right panel of \Fig{mapsOF_fid.fig}: they
are much denser, clumpier, (mostly) hotter, and more extended. Indeed,
looking at the dashed lines in \Fig{SFR_stoch.fig}, we see that the
outflow rate at $20$ kpc increases almost linearly with $\DTstoch$,
and is about $3$ times higher than the SFR (solid) at the end of the
run for $\DTstoch=10^9$ K. The average outflow velocity (not shown)
increases by a factor $2-3$ at $2$ kpc for the highest $\DTstoch$
considered (compared to the lowest), but is unaffected at $20$ kpc.

Varying $\DTstoch$ has a much weaker effect on the SFR than on the
outflow, as shown in \Fig{SFR_stoch.fig}. The `first' two increases in
$\DTstoch$ have almost no effect on the SFR, while the highest value
produces an initially higher SFR which then declines gradually, much
like in the case of kinetic feedback, and ends up significantly lower
than for lower $\DTstoch$ values. As with kinetic feedback, the
decline in SFR is likely due to the strong outflow depleting the
galaxy of fuel for star formation. Also, due to the lower heating
probability, star-forming clumps can continue to form stars longer
without having strong SN events disrupting them.

We recall that the SN energy is not directly redistributed to lower
gas densities with stochastic feedback (see Eq. \ref{pSN2.eq}). On the
contrary, we find that increasing $\DTstoch$ indirectly results in the
SN energy being deposited at higher gas densities due to the increased
clumpiness of the gas (not shown; $\DTstoch = 10^9$ K results in
energy being deposited at $\approx 0.5-1$ dex higher densities,
compared to the fiducial case). Hence the stochasticity increases
feedback efficiency purely by increasing the injected energy of a
given SN event \emph{at a given density} (while the total SN energy
over the galaxy is unchanged), and hence also the local cooling time.

In summary, increasing the stochasticity of feedback by increasing
$\DTstoch$ strongly increases the outflows. The SFRs are insensitive
to these stochasticity variations except for the highest value of
$\DTstoch$ considered. With very large stochasticity, the disc also
becomes increasingly clumpy.

\section{Varying the time-scale in delayed cooling
  feedback} \label{fb_dc.sec} Of the feedback models we have compared
in \Sec{comparison.sec}, with their fiducial setup parameters, delayed
cooling feedback suppresses star formation most strongly and yields
the highest mass loading factors for the outflows (see Figures
\ref{SFR.fig}, \ref{SFR_MW.fig}, \ref{KS_Fid.fig}, and
\ref{OFtime.fig}).

\begin{figure}
  \centering
  \includegraphics
    {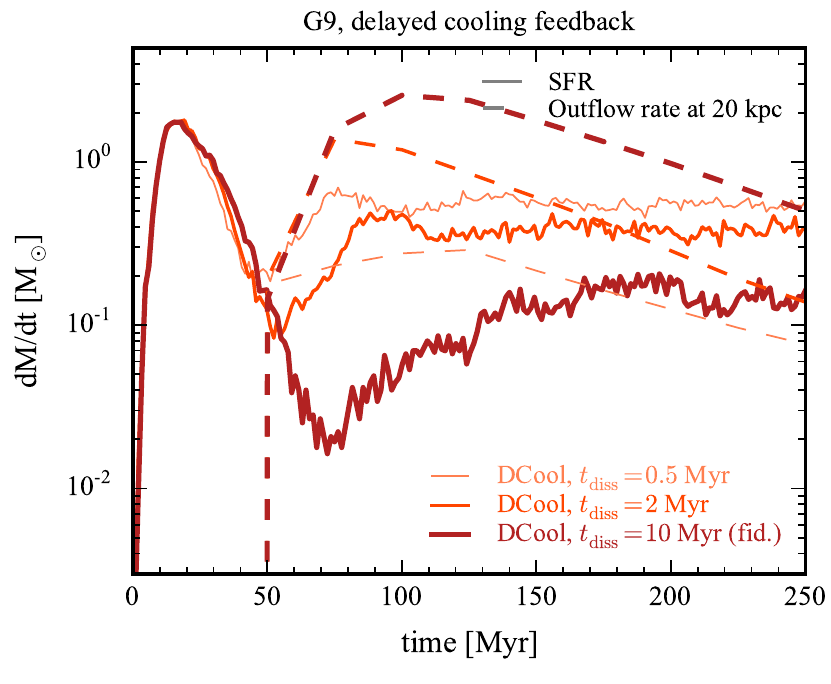}
  \caption
  {\label{SFR_dcool.fig}Star formation and gross outflow rates for the
    \sbc{} galaxy for variations in the delay time-scale, $\tDiss$, for
    delayed cooling feedback. The solid lines show SFRs, while dashed
    lines show outflow rates $20$ kpc from the disc. Decreasing
    $\tDiss$ from the fiducial value of $10$ Myr makes the SN feedback
    weaker. However, even at the lowest dissipation time-scale shown,
    the SFR at $250$ Myr is still low and the outflow rates are still
    high compared to the other feedback models (for their fiducial
    parameters).}
\end{figure}

Since delayed cooling feedback in its fiducial setup is strong, we
examine here what happens if we reduce the value of its free
parameter, which is the cooling delay time-scale, $\tDiss$. The
fiducial value is $\tDiss=10$ Myr, so here we compare with two runs
with significantly lower delay time-scales of $2$ and $0.5$ Myr. We
find that the results are highly sensitive to variations in
$\tDiss$. In \Fig{SFR_dcool.fig} we show the \sbc{} SFRs and gross
outflow rates across planes $20$ kpc from the disc, for those
variations. As expected, a shorter delay time-scale strongly decreases
the feedback efficiency, producing higher SFRs and lower outflow
rates. However, even for the shortest delay time-scale considered here,
the SN feedback is still more efficient in terms of suppressing star
formation than any of the other feedback models (and second only to
kinetic feedback in terms of outflow rates). Note that the mass
loading factor is particularly sensitive to the value of the free
parameter. At $250$ Myr it decreases from $\betaOut \approx 2$ for
$\tDiss =10$ Myr to $\betaOut \sim 10^{-1}$ for $\tDiss =0.5$ Myr
(both at $2$ and $20$ kpc from the disc). The outflow velocities,
which we do not show here, are unaffected at $20$ kpc, and are fairly
insensitive to $\tDiss$ at $2$ kpc (few tens of percent increase in
outflow velocity from smallest to highest $\tDiss$).

We analysed the KS relation for the same variations in $\tDiss$ (not
shown). Decreasing $\tDiss$ produces a KS relation which looks
increasingly like that of mechanical feedback (for the \sbc{} galaxy)
in \Fig{KS_Fid.fig}, with a similar slope and maximum gas surface
density of $7 \times 10^{2} \ \msunpc$ for $\tDiss=0.5$
Myr. Morphologically, the galaxy with the shortest dissipation
time-scale also looks quite similar (though a bit more diffuse) as the
runs with mechanical, stochastic, and thermal dump feedback in
\Fig{maps_Fid.fig}.

While the outflow rates decrease for shorter delay times-cales, the
morphological and phase structure of the outflows remain qualitatively
similar.

\section{Variations in kinetic SN feedback} \label{fb_kin.sec}

\begin{figure}
  \centering
  \includegraphics
    {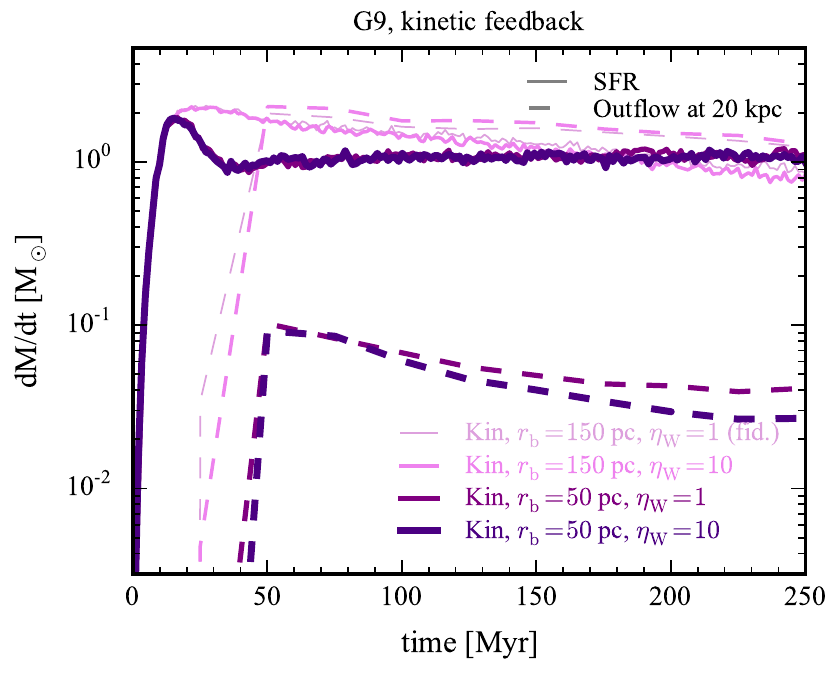}
  \caption
  {\label{SFR_kin.fig}Star formation and outflow rates for variations
    in the bubble radius ($\rb$) and sub-resolution mass loading
    ($\etaW$) for kinetic feedback in the \sbc{} galaxy. The solid
    lines show SFRs, while dashed lines show outflow rates $20$ kpc
    from the disc. Reducing the bubble radius (thicker curves) from
    the fiducial value of $150$ pc has a significant effect on the SFR
    and dramatically reduces the outflow rates. Changes in $\etaW$
    have little effect, but this is because the order unity ratio of
    the stellar particle mass to the local cell mass allows little
    room for increased sub-resolution mass loading.}
\end{figure}

In \Fig{SFR_kin.fig}, we show variations in the SFRs and gross mass
outflow rates at $20$ kpc for the \sbc{} galaxy, for variations in the
free kinetic feedback parameters, which are the bubble radius
(fiducially $150$ pc) and the sub-resolution mass loading parameter
$\etaW$ (fiducially set to $1$).

Decreasing the bubble radius, $\rb$, to about two times the minimum
cell width (of $18$ pc), has a significant effect on the SFR and
dramatically suppresses the gas outflow rate, which both (and also the
outflow speed) become similar to those produced by thermal dump,
stochastic, and mechanical feedback. Morphologically (not shown), the
runs with the small bubble radius also resemble the runs with these
other feedback models.

The smaller bubbles are less efficient at driving large-scale
outflows, because a larger fraction of the energy is now deposited
into dense ISM gas. For the same reason, the SFR at early times
decreases for smaller bubbles.  At late times the SFR is higher
because there remains substantially more gas in the galaxy as a
consequence of the weaker large-scale winds.

Increasing the sub-resolution mass loading, $\etaW$, by an order of
magnitude has little effect on star formation (\Fig{SFR_kin.fig}),
outflows (rates and speeds), and morphology (not shown). This is
because most SN events occur quite close to the density threshold for
star formation (see \Fig{SD_cum.fig}), meaning that usually there is
not much more mass available in the host cell than roughly matches the
stellar particle mass. Hence, an initial mass loading of more than
$\approx 1$ is not possible, since it would mean removing more mass
from the host cell than is available for re-distribution to the SN
bubble. To investigate the effect of lower stellar particle masses, or
star formation happening well above the density threshold, we
performed runs where we used a three times lower stellar particle
mass, $\mstar=600 \ \Msun$. Here increasing $\etaW$ \emph{reduces} the
outflow rates significantly (as it does to a smaller extent in
\Fig{SFR_kin.fig} for the smaller bubble radius). Moreover, the
feedback becomes more coupled to the disc, which becomes much
thicker. This can be understood from the fact that with a higher mass
loading, i.e. a larger ejected mass, the velocity of the ejected gas
must decrease to conserve momentum, and hence it is less likely to
escape from the galaxy.

We also looked at the KS relation for these variations in kinetic
feedback (not shown). Here, again, varying $\etaW$ has a negligible
effect, while the smaller bubble radius gives a relation very similar
to e.g. mechanical feedback.

\section{Discussion} \label{Discussion.sec}  

The effects of the SN feedback models we have studied are sensitive
to the resolution and/or mass of the simulated galaxies. For our
lower-mass \sbc{} galaxy, thermal dump, stochastic, and mechanical
feedback give very similar results in terms of SFRs, the
Kennicutt-Schmidt relation, and outflow mass loading factors, and we
have argued that this is an indication that the adiabatic phase of
thermal SN feedback is `resolved' in this galaxy, with the caveat that
we inject the equivalent of $40$ individual SN explosions
instantaneously. For our (ten times) more massive \mw{} galaxy,
however, this no longer applies, and thermal dump feedback is
significantly weaker than any of the other models.

If we compare our results to observations of SFRs in the local
universe, delayed cooling comes closest to reality. Observed SFRs for
local star-forming galaxies of similar stellar masses to \sbc{} are
typically in the range $\approx 0.05-0.4 \ \msunyr$ \citep[Figure 11
in][]{Chang2015}. In \Fig{SFR.fig}, we see that delayed cooling is
the only feedback model giving SFRs matching those observations,
while the other models give values well above the upper limits. For
stellar masses similar to our \mw{} galaxy ($10^{10} \, \Msun$),
\cite{Chang2015} give SFRs in the one-sigma range of
$\approx 0.3-1.5 \, \msunyr$, which actually falls slightly below the
delayed cooling SFR in \Fig{SFR_MW.fig}.  While such a comparison to
observations gives some indication of what is required to produce a
realistic suppression of star formation, we stress that comparing the
results from our isolated and somewhat idealised setup to observed
star formation rates has many caveats. Perhaps most significantly, the
gas fractions in our galaxies are high compared to those of local
galaxies \citep[see e.g. the compilations in][]{Bahe2016,
  Sales2016}.

Such caveats matter less if we consider the Kennicutt-Schmidt
relation. Here, all models except delayed cooling produce a slope
that is too steep compared to observations and/or SFR surface
densities that are significantly too high for a given gas surface
density. Delayed cooling does relatively well in terms of slope but
for our fiducial parameter choice it undershoots the SFR surface
density. However, unlike for the other models, this can be calibrated
towards a reasonable fit to the observed KS relation by tuning the
delayed cooling time-scale.

In addition, delayed cooling and `decoupled' kinetic feedback are the
only models able to produce mass loading factors exceeding unity,
while for other models with their fiducial parameters the mass
loading factors are at best an order of magnitude below unity.

While one can argue that these factors make delayed cooling for these
observables empirically most successful, the model produces
unrealistic thermal signatures in the ISM and CGM, where large amounts
of gas occupy a region in temperature-density space where the cooling
time is very short. Moreover, the convergence between thermal dump,
stochastic, and mechanical feedback suggest the adiabatic phase is
resolved and hence that the results from delayed cooling and kinetic
feedback may be unphysical.  

Mechanical feedback could be argued to be most physically motivated,
being based on analytic calculations and high (sub-pc) resolution
simulations of the final momentum attained in various environments in
terms of gas density and metallicity. Yet it does not come near
observations in suppressing star formation and produces weak outflows
in our simulations.  Perhaps this discrepancy can be traced to the
idealistic assumptions made when deriving the mechanical momentum
(Eq. \ref{dp_mech.eq}). While the momentum is realistic and converged
for a homogeneous medium, such a medium is not a good description of a
multi-phase and porous feedback-regulated star-forming ISM, and may
lead to an under-prediction of the generated momentum
\citep[e.g.][]{Kimm2015}.  We also neglect the preprocessing of the
local environment by stellar radiation, which lowers the surrounding
gas densities and has been shown to increase the momentum injection
from stellar populations, typically by a factor $\sim 2$
\citep[e.g.][]{Geen2015}.

It was recently reported by \cite{Gentry2016} that idealised
experiments of SN feedback in the literature have under-predicted the
final momentum by up to an order of magnitude, due to i) the neglect
of multiple successive SN explosions, ii) a lack of resolution, and
iii) a preference for Eulerian hydrodynamical solvers, which are
argued to suffer from over-mixing and hence over-cooling. If these
results are confirmed, we may have much more momentum to play with.

Another culprit may be the simplified setup of our simulations. For
control and to reduce the computational cost, we used isolated galaxy
simulations, assuming an initial state of the galaxy and its dark
matter halo rather than letting it evolve naturally in a cosmological
context. Ignoring environmental factors such as mergers and gas
accretion, may change the behaviour of SN feedback.

Finally, there are other feedback processes at play in galaxies that
we have neglected, such as AGN (thought to be important at high,
larger than $\approx$ MW masses; e.g. \citealt{Crain2015}), radiation
pressure \citep[e.g.][]{Rosdahl2015}, and cosmic rays
\citep[e.g.][]{Booth2013, Hanasz2013, Salem2014, Girichidis2016,
  Simpson2016}. The efficiency and interplay of each of those
processes is not well constrained, but they likely provide an
additional suppression of the SFR on top of SN feedback. If they turn
out to play an important role in galaxy evolution, some empirically
calibrated sub-resolution models for SN feedback may be re-interpreted
to also represent other feedback processes at play in galaxies.

\subsection{Dependence on factors other than SN feedback} \label{sfe.sec} 

Galaxy evolution involves a complex interplay of many physical
processes, and the SN feedback efficiencies that we have reported may
be sensitive to factors other than the SN feedback models.

One of the most important choices in our simulations aside from the
implementation of feedback is the local star formation efficiency,
which we have set to $\sfeff=2\%$. While this is a normal choice in
simulations of galaxy evolution, \cite{Agertz2015} argued that such
a low value artificially suppresses the effects of feedback, and that
higher local SF efficiencies of $\sfeff \approx 0.1$ are needed to
allow feedback to self-regulate star formation.  We have experimented
with $\sfeff=0.1$ in the \sbclr{} galaxy (while keeping the other
parameters fixed). While a full analysis is beyond the scope of the
present work, it is worth mentioning the qualitative effects. For all
SN feedback models considered, it results in a high early SFR peak,
followed by a dip and then convergence towards similar but somewhat
lower SFRs compared to those obtained for the fiducial
$\sfeff=0.02$. However, the SFR surface density moves in the wrong
direction, i.e. away from the observed KS relation, both in terms of
slope and normalisation.

We also investigated the effect of reducing $\sfeff$, to find what
calibration is required to reproduce the observed KS relation. A value
of $\sfeff=0.002$, i.e. ten times lower than our fiducial value,
produces a good match to the observed relation, but at the cost of
making the self-regulating effect of SN feedback negligible. In other
words, the $\sfeff$ parameter takes over as the dominant feedback
mechanism when set to a very low value.

While our non-comprehensive probes of the effect of a varying
numerical star formation efficiency $\sfeff$ are discouraging, we have
thus far not studied variations in $\sfeff$ in combination with other
factors. For example, in combination with a high local star formation
efficiency, early feedback may play a significant role by suppressing
runaway star formation for the $5$ Myr from the birth of the first
stars until the onset of SN explosions in a given star-forming region
\citep[e.g.][]{Hopkins2012b, Stinson2013, Agertz2015}. Also
the locality of star formation may be important for the efficiency of
feedback, i.e. it may matter whether the distribution of star
formation, and hence SN explosions, is scattered smoothly in space and
time, or happens in short localised bursts
\citep[e.g.][]{Federrath2013, Hopkins2013b, Walch2015a}.

We will investigate the role of $\sfeff$ in more detail in future
work, where we combine higher local SF efficiencies with more
stringent SF criteria and early feedback in the form of stellar
radiation.

There are many more variations which may or may not affect the
feedback efficiency, and again we have made limited explorations,
which we summarise below.
\begin{itemize}
\item We varied the density threshold for star formation, $\ns$, by a
  factor of ten in either direction, for stochastic feedback. Higher
  $\ns$ does give lower initial SFRs, but after $250$ Myr the values
  are the same to within a few percent for two orders of magnitude
  variations in $\ns$. Outflow rates remain nearly unchanged.
\item In the more massive galaxy we ran identical simulations with
  $0.1$ Solar metallicity, instead of the default Solar
  metallicity. The reduced metallicity has a marginal effect on the
  outflows and on the SFRs in the case of delayed cooling and kinetic
  feedback. However, for thermal dump, stochastic, and mechanical
  feedback, it reduces the SFRs by a few tens of percent and increases
  the outflow mass loading factor by about a factor of two.
\item Switching from non-equilibrium hydrogen and helium
  thermochemistry to an equilibrium assumption in the \sbc{} galaxy
  significantly increases outflow rates for thermal dump (factor
  $\sim 5$), stochastic feedback (factor $\sim 4$), and mechanical
  feedback (factor $\sim 5$), while the SFRs are only marginally
  reduced (and there is almost no effect on either SFRs or outflows
  with kinetic feedback or delayed cooling). In a forthcoming paper,
  we will explore the physical effects of equilibrium versus
  non-equilibrium thermochemistry in the context of SN feedback.
\item With thermal dump and stochastic feedback we scaled the Jeans
  pressure floor by a factor of three in each direction (i.e. three
  times higher and lower pressure at a given gas density). The disc
  morphology becomes slightly but noticeably smoother with a higher
  Jeans pressure, but the effect on star formation and outflows is
  negligible. A very large increase of the pressure floor suppresses
  the effect of SN feedback with any model, since it overtakes the
  role of SN feedback in suppressing the collapse of gas to
  star-forming densities.
\item With mechanical feedback, we explored the effect of runaway OB
  stars in the \sbc{} galaxy, giving a kick to each newborn stellar
  particle with a random direction and random speed of $0-50 \, \kms$.
  This has negligible effect on the SFRs, but the outflow rates are
  enhanced by almost a factor of ten, due to SN explosions taking
  place at significantly lower densities on average.
\item Also with mechanical feedback, we simulated individual $10^{51}$
  erg SN explosions, stochastically sampled for each stellar particle
  over the $5-50$ Myr lifetimes assumed for massive stars in the
  (Chabrier) IMF \citep[see details in][]{Kimm2015}. The effect is
  similar to the above test with runaway OB stars: the SFRs are
  marginally higher, while the outflow rates are increased
  significantly, though somewhat less than for the runaway OB
  stars. The reason for the increased outflow rates is that for a
  given particle, early SN explosions can clear away the dense
  environment, leading to late SN explosions taking place at reduced
  densities. We also combined runaway OB stars and individual SN
  explosions in the same run (again for the \sbc{} galaxy). This
  produces SFRs marginally higher than individual SNe only and
  outflows marginally higher than runaway OB stars only, i.e. it does
  not give an extra boost to the outflows.
\item Changing the time delay for stochastic SN feedback (zero to $20$
  Myr) has similar effects as runaway OB stars: The SFRs are only
  marginally affected, while an increased delay increases the outflow
  mass loading factor (it is halved for zero delay and doubles for a
  $20$ Myr delay).
\end{itemize}

\subsection{Comparison with stochastic heating in
  GADGET}\label{DS-comp.sec}
The stochastic feedback model included in this work is based on the
scheme introduced in \DS{} and used in the \gadget{} code. Similarly
to this work, \DS{} explore the effects of their stochastic feedback
model using rotating isolated galaxy discs of two masses, the main
difference being that their lower mass disc is roughly ten times less
massive than our \sbc{} disc (while their higher mass disc is
comparable in mass to our \mw{}). They compare different values for
stochastic heating and find $\DTstoch=10^{7.5}$ K to be efficient in
suppressing star formation (in the massive disc) and generating strong
outflows (in both discs). In \DS{} this value, which we also choose as
our fiducial value for stochastic heating, gives outflow mass loading
factors (at $20 \%$ of the virial radii) of $\approx 40$ and
$\approx 2$ for the low- and high-mass galaxies, respectively. This is
$\approx 40-200$ times larger than our mass loading factors, pointing
to a significant difference between our simulations and \DS{} in the
efficiency of stochastic feedback in suppressing star formation and
generating outflows.

In \App{ds.app} we discuss the differences between our simulations and
those of \DS{} and attempt to more closely reproduce their setup. We
conclude that there is a major difference between the two versions
(AMR and SPH) of stochastic feedback, with the SPH version being much
more efficient at generating outflows, for a given SFR. Pinpointing
the reason(s) for this difference, however, and whether it is due to
subtle differences in setup or resolution, or to more fundamental
differences between AMR and SPH, remains a challenge for follow-up
work.

\section{Conclusions} \label{Conclusions.sec} 

We used simulations of isolated galaxy discs with the \ramses{} code
to assess sub-resolution models for SN feedback in AMR simulations,
in particular their efficiency in suppressing star formation and
generating outflows. We focused our analysis on a dwarf galaxy, ten
times less massive than the MW, using a spatial resolution of $18$ pc
and a stellar (DM) mass resolution of $2 \times 10^3 \ \Msun$
($10^5 \ \Msun$), but also included a more limited analysis of a MW
mass galaxy (using a resolution of $36$ pc, $1.6 \times 10^4 \ \Msun$
stellar particles and $10^6 \ \Msun$ DM particles).

We studied five SN feedback models: i) thermal dump of SN energy into
the host cell of the star particle, ii) stochastic thermal feedback,
where the SN energy is re-distributed into fewer but more energetic
explosions, iii) kinetic feedback, where momentum is deposited
directly into a bubble around the star particle, iv) delayed cooling,
where cooling is suppressed temporarily in the expanding SN remnant,
and v) mechanical feedback, which injects energy or momentum depending
on the resolution. Three of those models can be calibrated with
adjustable parameters, which are the minimum local heating temperature
for stochastic feedback, the bubble size and local mass loading for
kinetic feedback, and the cooling suppression time for delayed
cooling). The mechanical feedback model has no free parameters (once
the SN energy has been decided) and the injected momentum is based on
analytic derivations and high-resolution simulations of cooling losses
in expanding SN blasts.  We compared the results produced using these
models with their fiducial settings, and for those models with
adjustable parameters we studied the effects of parameter
variations. Our main results are as follows.

For our low-mass, high-resolution galaxy, thermal dump, stochastic,
and mechanical feedback produce nearly identical results (Figs.
\ref{SFR.fig} and \ref{OFtime.fig}). We showed that at our current
resolution and star formation densities, stochastic feedback is
actually not that stochastic, and mechanical feedback is still mostly
in the adiabatic phase. Hence those feedback models are converged in
that setup, and thermal dump feedback adequately resolves the energy
injection (by multiple SNe in a single event). For our more massive
galaxy, stochastic and mechanical feedback become significantly
stronger than thermal dump feedback, but are still weak compared to
delayed cooling and kinetic feedback (\Fig{SFR_MW.fig}).

Strong outflows are not easily generated in our AMR simulations. Mass
loading factors of unity or above require extreme measures, such as
turning off cooling for a prolonged time, or kinetic feedback that is
in effect hydrodynamically decoupled due to the bubble radius
exceeding the disc height (\Fig{OFtime.fig}). The outflows produced by
delayed cooling and kinetic feedback are very distinct, the former
being cold, dense, and slow, while the latter are hot, diffuse, fast,
and featureless (Figs. \ref{SQO_Fid.fig}, \ref{mapsOF_fid.fig}, and
\ref{PH_Fid.fig}). The other models produce slow and remarkably
similar outflows at intermediate densities and temperatures.

Save for thermal dump feedback, all models do well in terms of
resolution convergence when considering SFRs, while, with the
exception of kinetic feedback, they produce significantly higher
outflow rates at lower resolution (\Fig{SFR_res.fig}).

Although a direct comparison is difficult, stochastic feedback appears
to produce much weaker outflows than in the similar disc runs with the
original SPH version of the model of \DS{}.  This discrepancy is
perhaps a result of subtle setup differences between our discs and
those of \DS{}, but we cannot rule out a more fundamental AMR versus
SPH difference. Stochastic feedback does become efficient at
generating massive outflows in our AMR discs if we use very high
values for the stochastic heating temperature (up to $10^9$ K), but
this comes at the cost of strong stochasticity due to low SN
probabilities (Figs. \ref{SFR_stoch.fig} and \ref{maps_stoch.fig}).

The major handle on the generation of outflows appears to be how well
the SN feedback model circumvents gas cooling, directly or
indirectly. Delayed cooling is the only model which succeeds at
generating outflows with mass loading factors exceeding unity
(\Fig{OFtime.fig}), at reproducing the observed main sequence SFRs
(Figs. \ref{SFR.fig} and \ref{SFR_MW.fig}), and the Kennicutt-Schmidt
relation (with appropriate calibration; \Fig{KS_Fid.fig}). The other
models fail to produce SN feedback strong enough to reproduce these
observations. This is discouraging, as delayed cooling retains too
much energy for too long, which in reality is partly lost to radiative
cooling, while the other feedback models are arguably more physically
motivated. Moreover, for the low-mass galaxy we argued that thermal
dump, stochastic, and mechanical feedback converge because we resolve
the adiabatic phase of the feedback events. This implies that in this
case delayed cooling and kinetic feedback yield incorrect answers.  In
particular, delayed cooling results in gas occupying regions of
temperature-density space where the cooling time is very short, which
compromises predictions for observational diagnostics.

Possible reasons for the disconnect between observations and our
results are: i) a lack of additional feedback physics, such as
radiation feedback or cosmic rays, ii) an incomplete setup, i.e. an
insufficiently realistic description of galaxies, iii) other aspects
of the subgrid physics, such as star formation, are unrealistic
\citep[e.g.][]{Agertz2015, Semenov2016}, iv) overcooling on
galactic scales is still an issue at our resolution, even if different
feedback models converge to the same results, and a significantly
higher resolution is required.

The current analysis will serve as a foundation for future studies of
feedback in galaxies, where we will use a sub-set of these models to
study the interplay of SN feedback with different sub-grid methods for
star formation and with feedback in the form of stellar radiation.

\section*{Acknowledgements}
We thank J\'er\'emy Blaizot, L\'eo Michel Dansac, Julien Devriendt,
Sylvia Ploeckinger, and Maxime Trebitsch for useful discussions, and
the anonymous referee for constructive comments. This work was funded
by the European Research Council under the European Union’s Seventh
Framework Programme (FP7/2007-2013) / ERC Grant agreement
278594-GasAroundGalaxies. TK was supported by the ERC Advanced Grant
320596 ``The Emergence of Structure during the Epoch of
Reionization''. The simulations were in part performed using the DiRAC
Data Centric system at Durham University, operated by the Institute
for Computational Cosmology on behalf of the STFC DiRAC HPC Facility
(www.dirac.ac.uk). This equipment was funded by BIS National
E-infrastructure capital grant ST/K00042X/1, STFC capital grant
ST/H008519/1, and STFC DiRAC Operations grant ST/K003267/1 and Durham
University. DiRAC is part of the National E-Infrastructure. We also
acknowledge PRACE for awarding us access to the ARCHER resource
(http://www.archer.ac.uk) based in the UK at the University of
Edinburgh (PRACE-3IP project FP7 RI-312763).

\bibliography{../../library}

\begin{thebibliography}{83}
\expandafter\ifx\csname natexlab\endcsname\relax\def\natexlab#1{#1}\fi

\bibitem[{Agertz \& Kravtsov(2015)}]{Agertz2015}
Agertz O., Kravtsov A.~V., 2015, AJ, 804, 18

\bibitem[{Agertz {et~al}\mbox{.}(2013)Agertz, Kravtsov, Leitner, \&
  Gnedin}]{Agertz2013}
Agertz O., Kravtsov A.~V., Leitner S.~N., Gnedin N.~Y., 2013, AJ, 770, 25

\bibitem[{Bah{\'{e}} {et~al}\mbox{.}(2016)Bah{\'{e}}, Crain, Kauffmann, Bower,
  Schaye, Furlong, Lagos, Schaller, Trayford, {Dalla Vecchia}, \&
  Theuns}]{Bahe2016}
Bah{\'{e}} Y.~M. {et~al.}, 2016, MNRAS, 456, 1115

\bibitem[{Barai {et~al}\mbox{.}(2015)Barai, Monaco, Murante, Ragagnin, \&
  Viel}]{Barai2015}
Barai P., Monaco P., Murante G., Ragagnin A., Viel M., 2015, MNRAS, 447, 266

\bibitem[{Bland-Hawthorn {et~al}\mbox{.}(2007)Bland-Hawthorn, Veilleux, \&
  Cecil}]{Bland-Hawthorn2007a}
Bland-Hawthorn J., Veilleux S., Cecil G., 2007, Astrophys. Space Sci., 311, 87

\bibitem[{Blondin {et~al}\mbox{.}(1998)Blondin, Wright, Borkowski, \&
  Reynolds}]{Blondin1998}
Blondin J.~M., Wright E.~B., Borkowski K.~J., Reynolds S.~P., 1998, AJ, 500,
  342

\bibitem[{Booth {et~al}\mbox{.}(2013)Booth, Agertz, Kravtsov, \&
  Gnedin}]{Booth2013}
Booth C.~M., Agertz O., Kravtsov A.~V., Gnedin N.~Y., 2013, ApJ, 777, L16

\bibitem[{Ceverino \& Klypin(2009)}]{Ceverino2009}
Ceverino D., Klypin A., 2009, AJ, 695, 292

\bibitem[{Chabrier(2003)}]{Chabrier2003}
Chabrier G., 2003, Publ. Astron. Soc. Pacific, 115, 763

\bibitem[{Chang {et~al}\mbox{.}(2015)Chang, van~der Wel, da~Cunha, \&
  Rix}]{Chang2015}
Chang Y.-Y., van~der Wel A., da~Cunha E., Rix H.-W., 2015, ApJS, 219, 8

\bibitem[{Crain {et~al}\mbox{.}(2015)Crain, Schaye, Bower, Furlong, Schaller,
  Theuns, {Dalla Vecchia}, Frenk, McCarthy, Helly, Jenkins, Rosas-Guevara,
  White, \& Trayford}]{Crain2015}
Crain R.~A. {et~al.}, 2015, MNRAS, 450, 1937

\bibitem[{Crighton {et~al}\mbox{.}(2013)Crighton, Hennawi, \&
  Prochaska}]{Crighton2013}
Crighton N. H.~M., Hennawi J.~F., Prochaska J.~X., 2013, ApJ, 776, L18

\bibitem[{{Dalla Vecchia} \& Schaye(2012)}]{DallaVecchia2012}
{Dalla Vecchia} C., Schaye J., 2012, MNRAS, 426, 140

\bibitem[{Dubois {et~al}\mbox{.}(2014)Dubois, Pichon, Welker, {Le Borgne},
  Devriendt, Laigle, Codis, Pogosyan, Arnouts, Benabed, Bertin, Blaizot,
  Bouchet, Cardoso, Colombi, de~Lapparent, Desjacques, Gavazzi, Kassin, Kimm,
  McCracken, Milliard, Peirani, Prunet, Rouberol, Silk, Slyz, Sousbie,
  Teyssier, Tresse, Treyer, Vibert, \& Volonteri}]{Dubois2014}
Dubois Y. {et~al.}, 2014, MNRAS, 444, 1453

\bibitem[{Dubois \& Teyssier(2008)}]{Dubois2008}
Dubois Y., Teyssier R., 2008, A{\&}A, 477, 79

\bibitem[{Dubois {et~al}\mbox{.}(2015)Dubois, Volonteri, Silk, Devriendt, Slyz,
  \& Teyssier}]{Dubois2015}
Dubois Y., Volonteri M., Silk J., Devriendt J., Slyz A., Teyssier R., 2015,
  MNRAS, 452, 1502

\bibitem[{Dwarkadas(2007)}]{Dwarkadas2007}
Dwarkadas V.~V., 2007, AJ, 667, 226

\bibitem[{Evans {et~al}\mbox{.}(2009)Evans, Dunham, J{\o}rgensen, Enoch,
  Mer{\'{i}}n, van Dishoeck, Alcal{\'{a}}, Myers, Stapelfeldt, Huard, Allen,
  Harvey, van Kempen, Blake, Koerner, Mundy, Padgett, \&
  Sargent}]{EvansNealJ2009}
Evans N.~J. {et~al.}, 2009, ApJS, 181, 321

\bibitem[{Faucher-Gigu{\`{e}}re {et~al}\mbox{.}(2009)Faucher-Gigu{\`{e}}re,
  Lidz, Zaldarriaga, \& Hernquist}]{Faucher-Giguere2009}
Faucher-Gigu{\`{e}}re C.-A., Lidz A., Zaldarriaga M., Hernquist L., 2009, AJ,
  703, 1416

\bibitem[{Federrath \& Klessen(2013)}]{Federrath2013}
Federrath C., Klessen R.~S., 2013, AJ, 763, 51

\bibitem[{Ferland {et~al}\mbox{.}(1998)Ferland, Korista, Verner, Ferguson,
  Kingdon, \& Verner}]{Ferland1998}
Ferland G.~J., Korista K.~T., Verner D.~A., Ferguson J.~W., Kingdon J.~B.,
  Verner E.~M., 1998, Publ. Astron. Soc. Pacific, 110, 761

\bibitem[{Fierlinger {et~al}\mbox{.}(2016)Fierlinger, Burkert, Ntormousi,
  Fierlinger, Schartmann, Ballone, Krause, \& Diehl}]{Fierlinger2016}
Fierlinger K.~M., Burkert A., Ntormousi E., Fierlinger P., Schartmann M.,
  Ballone A., Krause M. G.~H., Diehl R., 2016, MNRAS, 456, 710

\bibitem[{Geen {et~al}\mbox{.}(2015)Geen, Hennebelle, Tremblin, \&
  Rosdahl}]{Geen2015}
Geen S., Hennebelle P., Tremblin P., Rosdahl J., 2015, MNRAS, 454, 4484

\bibitem[{Gentry {et~al}\mbox{.}(2016)Gentry, Krumholz, Dekel, \&
  Madau}]{Gentry2016}
Gentry E.~S., Krumholz M.~R., Dekel A., Madau P., 2016, eprint arXiv:1606.01242

\bibitem[{Gerritsen(1997)}]{Gerritsen1997}
Gerritsen J. P.~E., 1997, Ph.D. thesis

\bibitem[{Gerritsen \& de~Blok(1999)}]{Gerritsen1999}
Gerritsen J. P.~E., de~Blok W. J.~G., 1999, A{\&}A, 342, 655

\bibitem[{Girichidis {et~al}\mbox{.}(2016)Girichidis, Naab, Walch, Hanasz, {Mac
  Low}, Ostriker, Gatto, Peters, W{\"{u}}nsch, Glover, Klessen, Clark, \&
  Baczynski}]{Girichidis2016}
Girichidis P. {et~al.}, 2016, ApJ, 816, L19

\bibitem[{Governato {et~al}\mbox{.}(2010)Governato, Brook, Mayer, Brooks, Rhee,
  Wadsley, Jonsson, Willman, Stinson, Quinn, \& Madau}]{Governato2010}
Governato F. {et~al.}, 2010, Nature, 463, 203

\bibitem[{Guillet \& Teyssier(2011)}]{Guillet2011}
Guillet T., Teyssier R., 2011, J. Comput. Phys., 230, 4756

\bibitem[{Haardt \& Madau(1996)}]{Haardt1996}
Haardt F., Madau P., 1996, AJ, 461, 20

\bibitem[{Haehnelt(1995)}]{Haehnelt1995}
Haehnelt M.~G., 1995, MNRAS, 273, 249

\bibitem[{Hanasz {et~al}\mbox{.}(2013)Hanasz, Lesch, Naab, Gawryszczak,
  Kowalik, \& W{\'{o}}lta{\'{n}}ski}]{Hanasz2013}
Hanasz M., Lesch H., Naab T., Gawryszczak A., Kowalik K., W{\'{o}}lta{\'{n}}ski
  D., 2013, ApJ, 777, L38

\bibitem[{Heckman {et~al}\mbox{.}(2015)Heckman, Alexandroff, Borthakur,
  Overzier, \& Leitherer}]{Heckman2015}
Heckman T.~M., Alexandroff R.~M., Borthakur S., Overzier R., Leitherer C.,
  2015, AJ, 809, 147

\bibitem[{Hopkins {et~al}\mbox{.}(2014)Hopkins, Kere{\v{s}}, O{\~{n}}orbe,
  Faucher-Gigu{\`{e}}re, Quataert, Murray, \& Bullock}]{Hopkins2014}
Hopkins P.~F., Kere{\v{s}} D., O{\~{n}}orbe J., Faucher-Gigu{\`{e}}re C.-A.,
  Quataert E., Murray N., Bullock J.~S., 2014, MNRAS, 445, 581

\bibitem[{Hopkins {et~al}\mbox{.}(2013)Hopkins, Narayanan, \&
  Murray}]{Hopkins2013b}
Hopkins P.~F., Narayanan D., Murray N., 2013, MNRAS, 432, 2647

\bibitem[{Hopkins {et~al}\mbox{.}(2012)Hopkins, Quataert, \&
  Murray}]{Hopkins2012b}
Hopkins P.~F., Quataert E., Murray N., 2012, MNRAS, 421, 3488

\bibitem[{Iffrig \& Hennebelle(2015)}]{Iffrig2015}
Iffrig O., Hennebelle P., 2015, A{\&}A, 576, A95

\bibitem[{Katz(1992)}]{Katz1992}
Katz N., 1992, AJ, 391, 502

\bibitem[{Keller {et~al}\mbox{.}(2014)Keller, Wadsley, Benincasa, \&
  Couchman}]{Keller2014}
Keller B.~W., Wadsley J., Benincasa S.~M., Couchman H. M.~P., 2014, MNRAS, 442,
  3013

\bibitem[{Kennicutt(1998)}]{Kennicutt1998}
Kennicutt R.~C., 1998, AJ, 498, 541

\bibitem[{Kim {et~al}\mbox{.}(2014)Kim, Abel, Agertz, Bryan, Ceverino,
  Christensen, Conroy, Dekel, Gnedin, Goldbaum, Guedes, Hahn, Hobbs, Hopkins,
  Hummels, Iannuzzi, Kere{\v{s}}, Klypin, Kravtsov, Krumholz, Kuhlen, Leitner,
  Madau, Mayer, Moody, Nagamine, Norman, O{\~{n}}orbe, O'Shea, Pillepich,
  Primack, Quinn, Read, Robertson, Rocha, Rudd, Shen, Smith, Szalay, Teyssier,
  Thompson, Todoroki, Turk, Wadsley, Wise, Zolotov, \& the {AGORA
  Collaboration29}}]{Kim2014}
Kim J.-h. {et~al.}, 2014, ApJS, 210, 14

\bibitem[{Kimm \& Cen(2014)}]{Kimm2014}
Kimm T., Cen R., 2014, AJ, 788, 121

\bibitem[{Kimm {et~al}\mbox{.}(2015)Kimm, Cen, Devriendt, Dubois, \&
  Slyz}]{Kimm2015}
Kimm T., Cen R., Devriendt J., Dubois Y., Slyz A., 2015, MNRAS, 451, 2900

\bibitem[{Krumholz \& Tan(2007)}]{Krumholz2007}
Krumholz M.~R., Tan J.~C., 2007, AJ, 654, 304

\bibitem[{Macci{\`{o}} {et~al}\mbox{.}(2008)Macci{\`{o}}, Dutton, \& van~den
  Bosch}]{Maccio2008}
Macci{\`{o}} A.~V., Dutton A.~A., van~den Bosch F.~C., 2008, MNRAS, 391, 1940

\bibitem[{Martizzi {et~al}\mbox{.}(2015)Martizzi, Faucher-Gigu{\`{e}}re, \&
  Quataert}]{Martizzi2015}
Martizzi D., Faucher-Gigu{\`{e}}re C.-A., Quataert E., 2015, MNRAS, 450, 504

\bibitem[{Mitra {et~al}\mbox{.}(2015)Mitra, Dav{\'{e}}, \&
  Finlator}]{Mitra2015}
Mitra S., Dav{\'{e}} R., Finlator K., 2015, MNRAS, 452, 1184

\bibitem[{Mo \& Mao(2004)}]{Mo2004}
Mo H.~J., Mao S., 2004, MNRAS, 353, 829

\bibitem[{Mollitor {et~al}\mbox{.}(2015)Mollitor, Nezri, \&
  Teyssier}]{Mollitor2015}
Mollitor P., Nezri E., Teyssier R., 2015, MNRAS, 447, 1353

\bibitem[{Murray {et~al}\mbox{.}(2010)Murray, Quataert, \&
  Thompson}]{Murray2010}
Murray N., Quataert E., Thompson T.~A., 2010, AJ, 709, 191

\bibitem[{Navarro {et~al}\mbox{.}(1997)Navarro, Frenk, \& White}]{Navarro1997}
Navarro J.~F., Frenk C.~S., White S. D.~M., 1997, AJ, 490, 493

\bibitem[{Navarro \& White(1993)}]{Navarro1993}
Navarro J.~F., White S. D.~M., 1993, MNRAS, 265, 271

\bibitem[{Puchwein \& Springel(2013)}]{Puchwein2013}
Puchwein E., Springel V., 2013, MNRAS, 428, 2966

\bibitem[{Rasera \& Teyssier(2006)}]{Rasera2006}
Rasera Y., Teyssier R., 2006, A{\&}A, 445, 1

\bibitem[{Rieder \& Teyssier(2016)}]{Rieder2016}
Rieder M., Teyssier R., 2016, MNRAS, 457, 1722

\bibitem[{Rogers \& Pittard(2013)}]{Rogers2013}
Rogers H., Pittard J.~M., 2013, MNRAS, 431, 1337

\bibitem[{Rosdahl {et~al}\mbox{.}(2013)Rosdahl, Blaizot, Aubert, Stranex, \&
  Teyssier}]{Rosdahl2013}
Rosdahl J., Blaizot J., Aubert D., Stranex T., Teyssier R., 2013, MNRAS, 436,
  2188

\bibitem[{Rosdahl {et~al}\mbox{.}(2015)Rosdahl, Schaye, Teyssier, \&
  Agertz}]{Rosdahl2015}
Rosdahl J., Schaye J., Teyssier R., Agertz O., 2015, MNRAS, 451, 4553

\bibitem[{Rosen \& Bregman(1995)}]{Rosen1995}
Rosen A., Bregman J.~N., 1995, AJ, 440, 634

\bibitem[{Ro{\v{s}}kar {et~al}\mbox{.}(2014)Ro{\v{s}}kar, Teyssier, Agertz,
  Wetzstein, \& Moore}]{Roskar2014}
Ro{\v{s}}kar R., Teyssier R., Agertz O., Wetzstein M., Moore B., 2014, MNRAS,
  444, 2837

\bibitem[{Salem \& Bryan(2014)}]{Salem2014}
Salem M., Bryan G.~L., 2014, MNRAS, 437, 3312

\bibitem[{Sales {et~al}\mbox{.}(2016)Sales, Navarro, Oman, Fattahi, Ferrero,
  Abadi, Bower, Crain, Frenk, Sawala, Schaller, Schaye, Theuns, \&
  White}]{Sales2016}
Sales L.~V. {et~al.}, 2016, MNRAS, 464, 2419

\bibitem[{Schaye {et~al}\mbox{.}(2015)Schaye, Crain, Bower, Furlong, Schaller,
  Theuns, {Dalla Vecchia}, Frenk, McCarthy, Helly, Jenkins, Rosas-Guevara,
  White, Baes, Booth, Camps, Navarro, Qu, Rahmati, Sawala, Thomas, \&
  Trayford}]{Schaye2015}
Schaye J. {et~al.}, 2015, MNRAS, 446, 521

\bibitem[{Schroetter {et~al}\mbox{.}(2015)Schroetter, Bouch{\'{e}}, Peroux,
  Murphy, Contini, \& Finley}]{Schroetter2015}
Schroetter I., Bouch{\'{e}} N., Peroux C., Murphy M.~T., Contini T., Finley H.,
  2015, AJ, 804, 83

\bibitem[{Semenov {et~al}\mbox{.}(2016)Semenov, Kravtsov, Gnedin, V., A.,
  {Kravtsov A. V.}, O., {Lake G.}, {Agertz O.}, B., F., C., {Leroy A.}, {Leroy
  A. K.}, {Rosolowsky E.}, K., {Teyssier R.}, G., W., Schmidt, W., Schmidt,
  {Schmidt W.}, H., {Stone J. M.}, L., P., J., A., Klypin, S., {Bonnell I. A.},
  C., M., H., {D'Onghia E.}, L., {Dekel A.}, D., {Dobbs C. L.}, E., {Evans N.
  J.II}, N., C., S., R., {Schr{\"{o}}n M.}, C., C., E., {Gammie C. F.}, J.,
  {Garnier E.}, P., {Naab T.}, {Gnedin N. Y.}, V., {Brook C.}, {Allen L. E.},
  J.II, G., Chabrier, L., {Kere{\v{s}} D.}, {Hopkins P. F.}, N., {Hopkins P.
  F.}, N., {Schmidt W.}, L., L., Dedes, N., J., N., C.Jr., A., S., Basu, {Abel
  T.}, V., {Kravtsov A. V.}, Y., {Kravtsov A. V.}, M., {Kritsuk A. G.}, L.,
  {Nordlund {\AA}.}, {Krumholz M. R.}, F., {Krumholz M. R.}, F., F., C., {Lada
  C. J.}, F., B., {Schleicher D. R. G.}, A., {Toschi F.}, E., {Dolphin A. E.},
  {Klessen R. S.}, M.-M., {Machacek M. E.}, T., {Iapichino L.}, A., {Martizzi
  D.}, E., {Schinnerer E.}, M., J., {Papamastorakis J.}, M., N., Y., Sofue,
  {Navarro J. F.}, M., {Padoan P.}, {\AA}., {\AA}., Nordlund, P., M., R., M.,
  {Wong T.}, P., Goldreich, V., A., {Hillebrandt W.}, {Rudd D. H.}, V., P., M.,
  M., W., {Almgren A. S.}, C., Federrath, {Seth A.}, {Brook C.}, {Stone J. M.},
  F., {Rix H.-W.}, {Klein R. I.}, B., {Li Z.-Y.}, P., P., \& Y.}]{Semenov2016}
Semenov V.~A. {et~al.}, 2016, AJ, 826, 200

\bibitem[{Simpson {et~al}\mbox{.}(2016)Simpson, Pakmor, Marinacci, Pfrommer,
  Springel, Glover, Clark, \& Smith}]{Simpson2016}
Simpson C.~M., Pakmor R., Marinacci F., Pfrommer C., Springel V., Glover S.
  C.~O., Clark P.~C., Smith R.~J., 2016, AJ, 827, L29

\bibitem[{Somerville \& Dav{\'{e}}(2015)}]{Somerville2015}
Somerville R.~S., Dav{\'{e}} R., 2015, Annu. Rev. Astron. Astrophys., 53, 51

\bibitem[{Springel {et~al}\mbox{.}(2005)Springel, {Di Matteo}, \&
  Hernquist}]{Springel2005a}
Springel V., {Di Matteo} T., Hernquist L., 2005, MNRAS, 361, 776

\bibitem[{Springel \& Hernquist(2003)}]{Springel2003}
Springel V., Hernquist L., 2003, Astrophys. Supercomput. using Part.
  Simulations, 208, 273

\bibitem[{Steidel {et~al}\mbox{.}(2010)Steidel, Erb, Shapley, Pettini, Reddy,
  Bogosavljevi{\'{c}}, Rudie, \& Rakic}]{Steidel2010}
Steidel C.~C., Erb D.~K., Shapley A.~E., Pettini M., Reddy N.~A.,
  Bogosavljevi{\'{c}} M., Rudie G.~C., Rakic O., 2010, AJ, 717, 289

\bibitem[{Stinson {et~al}\mbox{.}(2006)Stinson, Seth, Katz, Wadsley, Governato,
  \& Quinn}]{Stinson2006}
Stinson G., Seth A., Katz N., Wadsley J., Governato F., Quinn T., 2006, MNRAS,
  373, 1074

\bibitem[{Stinson {et~al}\mbox{.}(2013)Stinson, Brook, Macci{\`{o}}, Wadsley,
  Quinn, \& Couchman}]{Stinson2013}
Stinson G.~S., Brook C., Macci{\`{o}} A.~V., Wadsley J., Quinn T.~R., Couchman
  H. M.~P., 2013, MNRAS, 428, 129

\bibitem[{Teyssier(2002)}]{Teyssier2002}
Teyssier R., 2002, A{\&}A, 385, 337

\bibitem[{Teyssier {et~al}\mbox{.}(2013)Teyssier, Pontzen, Dubois, \&
  Read}]{Teyssier2013}
Teyssier R., Pontzen A., Dubois Y., Read J.~I., 2013, MNRAS, 429, 3068

\bibitem[{Thompson {et~al}\mbox{.}(2005)Thompson, Quataert, \&
  Murray}]{Thompson2005}
Thompson T.~A., Quataert E., Murray N., 2005, AJ, 630, 167

\bibitem[{Thornton {et~al}\mbox{.}(1998)Thornton, Gaudlitz, Janka, \&
  Steinmetz}]{Thornton1998}
Thornton K., Gaudlitz M., Janka H.-T., Steinmetz M., 1998, AJ, 500, 95

\bibitem[{Toro {et~al}\mbox{.}(1994)Toro, Spruce, \& Speares}]{Toro1994}
Toro E.~F., Spruce M., Speares W., 1994, Shock Waves, 4, 25

\bibitem[{Truelove {et~al}\mbox{.}(1997)Truelove, Klein, McKee, Holliman,
  Howell, \& Greenough}]{Truelove1997}
Truelove J.~K., Klein R.~I., McKee C.~F., Holliman J.~H., Howell L.~H.,
  Greenough J.~A., 1997, AJ, 489, L179

\bibitem[{Veilleux {et~al}\mbox{.}(2005)Veilleux, Cecil, \&
  Bland-Hawthorn}]{Veilleux2005}
Veilleux S., Cecil G., Bland-Hawthorn J., 2005, Annu. Rev. Astron. Astrophys.,
  43, 769

\bibitem[{Vogelsberger {et~al}\mbox{.}(2013)Vogelsberger, Genel, Sijacki,
  Torrey, Springel, \& Hernquist}]{Vogelsberger2013}
Vogelsberger M., Genel S., Sijacki D., Torrey P., Springel V., Hernquist L.,
  2013, MNRAS, 436, 3031

\bibitem[{Walch {et~al}\mbox{.}(2015)Walch, Girichidis, Naab, Gatto, Glover,
  W{\"{u}}nsch, Klessen, Clark, Peters, Derigs, \& Baczynski}]{Walch2015a}
Walch S. {et~al.}, 2015, MNRAS, 454, 238

\bibitem[{Wang {et~al}\mbox{.}(2015)Wang, Dutton, Stinson, Macci{\`{o}}, Penzo,
  Kang, Keller, \& Wadsley}]{Wang2015}
Wang L., Dutton A.~A., Stinson G.~S., Macci{\`{o}} A.~V., Penzo C., Kang X.,
  Keller B.~W., Wadsley J., 2015, MNRAS, 454, 83

\bibitem[{Zuckerman \& Evans(1974)}]{Zuckerman1974}
Zuckerman B., Evans N.~J., 1974, AJ, 192, L149

\end{thebibliography}


\appendix

\section{Stochastic feedback comparison with \DS{}} \label{ds.app}

\DS{} introduced a model for stochastic SN feedback in the SPH code
\gadget{}, on which we base our AMR version of stochastic feedback. In
simulations of two isolated disc galaxies and using the same fiducial
stochastic heating temperature difference as us of $\DTstoch=10^{7.5}$
K, they find mass loading factors of $\betaOut\approx40$ and
$\betaOut\approx2$, at $20\%$ of the virial radius, for galaxies of
baryonic masses $4 \times 10^8 \, \Msun$ and
$4 \times 10^{10} \, \Msun$, respectively. For a lower-mass galaxy
roughly ten times lower in mass than theirs and a higher-mass galaxy
roughly similar to theirs, we find mass loading factors at comparable
distances that are $1-2$ orders of magnitude lower than their values.

While this may be due to the different hydrodynamical solvers,
i.e. SPH and AMR, a direct comparison is complicated by the fact that
in a number of ways, our simulations are set up differently from those
in \DS{}.

One difference is that in the disc simulations of \DS{}, the CGM
initially has zero gas density, while we are forced in AMR
hydrodynamics to use positive nonzero density everywhere. We can rule
this out as a major issue though: we ran simulations where we changed
the initial CGM density by factors of ten in each direction, and found
our simulation results remain unchanged.

The remaining trivial setup differences are the different pressure
floors, cooling functions, star formation criteria, and time delay
between the birth of stellar particles and their SN events. To assess
the significance of those differences, we have performed a \mw{} run,
and also of a galaxy similar to the low-mass galaxy in \DS{}, with our
fiducial stochastic feedback, but with a setup closer to that in
\DS{}, i.e. with zero metallicity, equilibrium cooling,
$\ns=0.1 \ \cci$, an (almost) identical pressure floor,
\begin{displaymath}
\TJeans=10^4 \ {\rm K} \left( \frac{\nh}{0.1 \ \cci} \right)^{1/3},
\end{displaymath}
and a time delay of $30$ Myr from the birth and SN events of stellar
particles (a factor of $6$ longer than our fiducial time delay).

Although this brings us closer to the simulation settings of \DS{}, we
emphasize that it is still not an ideal comparison. For one thing, the
pressure floor is applied slightly differently. In \DS{} it is an
actual floor, while in our simulations, it is added to the thermal
pressure which is evolved separately. Also, in \DS{} the pressure
floor is only applied above the density threshold for star formation,
of $\nh=0.1 \ \cci$, while we apply it everywhere (though in the
absence of metals, gas does not cool below the floor below the density
threshold for star formation). Third, the star formation law in \DS{}
is pressure-dependent, and hence very different from ours. Fourth, it
is not obvious whether the resolution is comparable between the AMR
and SPH runs\footnote{While the number and mass of DM particles and
  gravitational `softening' scales are similar between our simulations
  and those of DS{} are comparable, it becomes more tricky for both
  the gas mass and physical resolution around SN events. In our AMR
  runs, the mass resolution of the gas elements receiving the SN
  energy is highly variable and depends linearly on the gas density,
  while in SPH the gas (and stellar) mass resolution is fixed
  ($5.1\times10^2 \, \Msun$ and $5.1\times10^4 \, \Msun$ for the
  low-mass and high-mass galaxies in \DS{}, respectively). At the same
  time, the physical resolution is fixed in AMR (assuming the highest
  refinement level), but variable in SPH (depending, again, on the
  density).} Still, we find no reason to expect that any of these
discrepancies should lead to order of magnitude differences in the
mass loading factors.

\begin{figure}
  \centering
  \includegraphics
    {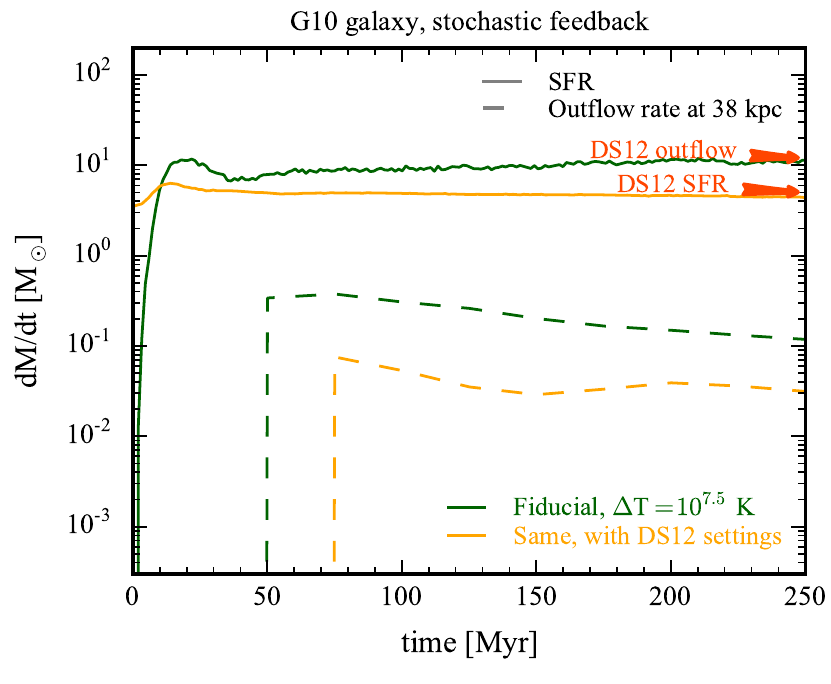}
  \caption
  {\label{SFRcomp_g10.fig}Approximate comparison with \DS{} of SFRs
    (solid curves) and outflow rates (dashed). Green (dark) curves are
    from a run with our fiducial settings, as presented in the bulk of
    the paper, while orange (lighter) curves are using settings more
    similar to those in \DS{} (metallicity, pressure floor, SF density
    threshold). Red arrows indicate the SFR and outflow rate at $250$
    Myr in the corresponding \DS{} run (their {\sc g12}), where the
    SFR is similar to ours (but steadily declining), but the outflow
    rate, and hence the mass loading factor, is more than two orders
    of magnitude higher than in our runs.}
\end{figure}

In \Fig{SFRcomp_g10.fig}, we show the SFRs and outflow rates at
$20 \%$ of the virial radius for our massive \mw{} galaxy and
stochastic feedback, both for our fiducial simulation settings (green
curves) and for these alternative settings described above to mimic
those of \DS{} (orange curves). With arrows at the right side of the
plot we indicate the SFR and outflow rate at $250$ Myr in the
corresponding run in \DS{} (their {\sc g12} galaxy), also with
stochastic feedback and $\DTstoch=10^{7.5}$ K. By comparing the green
and orange solid curves we see that applying the \DS{} simulation
settings has the effect of reducing the SFR by a factor $\approx 2$
and reducing the outflow rate by a factor of few. Clearly, this does
not help in bringing the mass loading factor closer to the one found
in \DS{}. The SFR in our run mimicking the \DS{} simulation settings
is actually close to that in \DS{}, and the suppression is likely
mostly due to the strong pressure floor. However, even with a similar
SFR as in \DS{}, the outflow rate is more than two orders of magnitude
lower.

\begin{figure}
  \centering
  \includegraphics
    {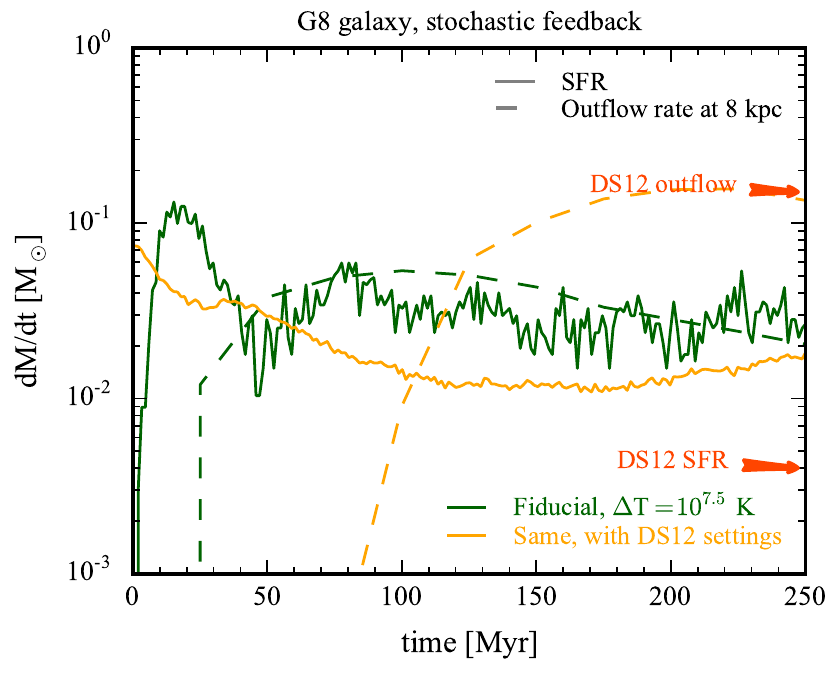}
  \caption
  {\label{SFRcomp_g8.fig}Same as \Fig{SFRcomp_g10.fig}, but for the
    `super-dwarf' \dw{} galaxy, which is similar to the low-mass
    galaxy in \DS{} (their {\sc g10}). The SFR in the \DS{} run (red
    arrow) is much lower than in our fiducial run (green), and
    conversely the outflow rate much higher, leading to more than an
    order of magnitude difference in the mass loading factor. For the
    run where we attempt to mimic the settings of \DS{} (orange
    curves), the outflow rate is similar to \DS{}, and hence the mass
    loading factor becomes closer to that of \DS{}, although it is
    still a factor of few lower due to the higher SFR.}
\end{figure}

\Fig{SFRcomp_g8.fig} shows the same kind of comparison, but now for a
galaxy with a baryonic mass of $3.5 \times 10^8 \ \Msun$, similar to
the low-mass galaxy in \DS{}. We use the exact setup parameters as for
the lowest-mass \dw{} galaxy described in \cite{Rosdahl2015},
except that as with the other galaxies in this paper, the mass of
formed stellar particles is $\ns{}$ times the cell volume at maximum
resolution, and we now include a UV background. We performed one run
with the fiducial settings described in \Sec{simulations.sec}, and
another run with the \DS{} settings described above for the \mw{}
galaxy.

Comparing to the two arrows from \DS{}, our SFRs are a nearly an order
of magnitude higher and not highly sensitive to the setup, except it
is more bursty with our fiducial setup. The outflow rate, however is
quite sensitive to the setup, being nearly an order of magnitude lower
than that of \DS{} for our fiducial setup, but nearly identical to
that of \DS{} when we mimic their setup. Hence, the mass loading
factor is more than an order of magnitude lower in our fiducial runs
and a factor of $4-5$ lower in our runs mimicking the \DS{}
settings. As with the \mw{} galaxy, though not as significantly, SN
feedback is less efficient at regulating our galaxy and generating
outflows than in \DS{}, hinting towards subtle `non-linear'
differences in simulation settings that remain undetected, or more
fundamental differences between the hydrodynamical solvers (or
both). We stress, however, that we must treat these hints with
caution. While differences between SPH and AMR in terms of feedback
efficiency and the generation of winds may exist, we cannot rule out
other causes for now, and a further exploration must be left for
future work.

\end{document}